\documentclass[prx,twocolumn,english,superscriptaddress,floatfix,longbibliography,nofootinbib]{revtex4-2}

\usepackage{style}
\usepackage{mathrsfs}
\usepackage{amsmath}

\usepackage{graphicx}
\usepackage{dcolumn}
\usepackage{bm}
\usepackage{gensymb}
\usepackage{wrapfig}
\usepackage{natbib}
\usepackage{float}
\usepackage{titlesec}
\usepackage{comment}
\usepackage{xcolor}
\usepackage{makecell}
\usepackage{caption}
\begin{document}

\newcommand{\Pb}{\mathrm{Pb}}
\newcommand{\Po}{\mathrm{Po}}
\newcommand{\Bi}{\mathrm{Bi}}


\title{Accelerating Surface Radiation Content to Investigate the Impact of Radon Progeny on Superconducting Qubits}

\author{Sagar S. Poudel}
\affiliation{Physics Department, South Dakota School of Mines \& Technology, Rapid City, SD, USA 57701}
\author{Dylan J. Temples}\thanks{Contact email: \href{mailto:dtemples@fnal.gov}{dtemples@fnal.gov}}
\affiliation{Fermi National Accelerator Laboratory, Batavia, IL, USA 60510}
\author{Ryan Linehan}
\affiliation{Fermi National Accelerator Laboratory, Batavia, IL, USA 60510}
\affiliation{Department of Physics \& Astronomy, Northwestern University, Evanston, IL, USA 60208}
\author{Alejandro Rodriguez}
\affiliation{Department of Physics \& Astronomy, Northwestern University, Evanston, IL, USA 60208}
\author{Matthew Hall}
\affiliation{Department of Mathematics, University of Southern California, Los Angeles, CA, USA 90089}
\affiliation{Fermi National Accelerator Laboratory, Batavia, IL, USA 60510}
\author{Dax Kay}
\affiliation{Department of Physics \& Astronomy, Northwestern University, Evanston, IL, USA 60208}
\affiliation{Fermi National Accelerator Laboratory, Batavia, IL, USA 60510}
\author{Nathaniel Rodenburg}
\affiliation{Physics Department, South Dakota School of Mines \& Technology, Rapid City, SD, USA 57701}
\author{Daniel Baxter}
\affiliation{Fermi National Accelerator Laboratory, Batavia, IL, USA 60510}
\affiliation{Department of Physics \& Astronomy, Northwestern University, Evanston, IL, USA 60208}
\author{Enectali Figueroa-Feliciano}
\affiliation{Department of Physics \& Astronomy, Northwestern University, Evanston, IL, USA 60208}
\affiliation{Fermi National Accelerator Laboratory, Batavia, IL, USA 60510}
\author{Richard W. Schnee}
\affiliation{Physics Department, South Dakota School of Mines \& Technology, Rapid City, SD, USA 57701}
\author{Lauren Hsu}
\affiliation{Fermi National Accelerator Laboratory, Batavia, IL, USA 60510}

\date{May 29, 2026}

\begin{abstract}

Ionizing radiation in the form of $\alpha$, $\beta$, $\gamma$, and additional high-energy particles can induce decoherence via phonon and quasiparticle poisoning in superconducting qubits. Recent studies have explored this effect using cosmic rays or controlled radioactive sources held in the proximity of a qubit package, and have concluded that reductions in such ``external'' environmental radiation may benefit stable operation of qubit devices. However, the effect of long-lived, unstable daughters of $^{222}$Rn that ``plate out'' directly on device and packaging surfaces has not been as extensively explored. This plate-out process, well-known to the dark matter direct detection field, occurs throughout the fabrication and testing lifecycle of a device and (separately) its packaging, and produces a local source of $\alpha$-decays which can remain active for decades. As this scales with chip area, understanding and managing this source of ionizing radiation is relevant for successfully scaling quantum computing architectures to larger numbers of qubits in a radiation-robust way. We present a setup capable of accelerating and enhancing radon daughter plateout by a factor of $7\times10^4$ over ambient, in order to study, \textit{in situ}, the impact of these events on superconducting qubits. We also provide outlook on the potential impact of this source of ionizing radiation on current and future qubit arrays.\\
\textbf{FNAL Report Number:} FERMILAB-PUB-26-0276-ETD-PPD
\end{abstract}
\maketitle

\section{Introduction}

In recent decades, superconducting circuit platforms have arisen as a promising candidate for construction of a fault-tolerant quantum computer \cite{Nakamura1999,StateOfPlay2020}. These platforms typically consist of an array of superconducting qubits etched into a thin film superconducting material such as aluminum, niobium, or tantalum, deposited on top of thicker substrates such as silicon or sapphire~\cite{Oliver2013,Murray2021}. Superconducting qubits must retain the coherence of fragile electromagnetic quantum states to perform noise-free computations. This requirement has spurred effort to improve various qubit coherence metrics: the deexcitation time and dephasing time ($T_1$ and $T_2$, respectively), via advances in materials and fabrication~\cite{Bal2024,Bland2025,Bland2025a,Iaia2022}, infrared (IR) shielding \cite{Barends2011, Corcoles2011, Pan2022, Liu2024, Kerschbaum2026}, and qubit design~\cite{Koch2007,Gambetta2017,Martinis2022,McEwen2024,Tuokkola2025,Pinckney2026}. In the last decade, on-chip radiation impacts have been identified and explored as a source of qubit coherence degradation~\cite{Martinis2021, Vepsalainen:2020trd,Larson2025,McEwen2021,Harrington2025,Wilen,Tennant2022,Bratrud:2024qnk,Thorbeck:2022yzs,Pinckney2026}.  These events have been identified as causing spatiotemporally correlated bit-flip errors~\cite{McEwen2021, Harrington2025} and dephasing errors~\cite{Wilen,Tennant2022,Bratrud:2024qnk}, both of which pose a problem for surface-code error-correction algorithms~\cite{Dennis2002,Fowler2012,Nickerson2019,Acharya2024,Wang2025b,Tan2026} which expect qubit errors to be uncorrelated. Furthermore, there is evidence that this radiation may lead to scrambling of the electrically-active two-level systems (TLS) that act as part of qubits' dissipative baths~\cite{Thorbeck:2022yzs}. 

Superconducting qubits' susceptibility to radiation impacts is problematic for fault-tolerant computing in part because a total suppression of ambient radiation backgrounds is impossible. Cosmic rays (CRs) and $\gamma$ rays from ambient radioactive decays will yield a baseline rate of scatters in a chip in a surface lab~\cite{Harrington2025,Fowler2024}. While qubit and chip design can be tuned to reduce the creation of correlated errors during these impacts~\cite{Downconversion, McEwen2024,kurilovich2025correlatederrorburstsgapengineered}, such strategies have not yet offered complete protection against high-energy naturally-occurring backgrounds~\cite{Pinckney2026,kurilovich2025correlatederrorburstsgapengineered}. While CR muons and $\gamma$-rays from ambient radioactivity are the most common background and deposit $\mathcal{O}(10-1000)$~keV of energy into a standard wafer (which may be easily mitigated), other sources of radiation, including CR protons and neutrons, occur more rarely~\cite{PDG2024CosmicRays} but with far more energy deposited, ranging up to several MeV or higher~\cite{Fowler2024}. Shielding and underground operation may further reduce these backgrounds~\cite{Bratrud:2024qnk,Cardani2023Disentangling}, but is challenging to scale for commercial applications.

Even with mitigation of the above radiation sources, another source of high-energy radiation will exist: $\alpha$ decays ($\approx$5~MeV).  A common source arises from radon daughters plated out onto a device or its packaging~\cite{LinehanLZRadon2026,Chernyak2023}. This plate-out may occur directly onto a chip or inward-facing surfaces of its enclosure during device fabrication and packaging, giving the source of the alpha decays a direct line-of-sight to the device. While the total rate of these decays is fairly low for small, demonstrator-style chips ($\approx$0.1 mHz/cm$^{2}$), they are impossible to mitigate through shielding and are challenging to mitigate through external vetoes due to the short range of $\alpha$'s in matter.  Their prevalence is also highly dependent on the environmental history of the qubits, meaning that reduction of this background requires deliberate efforts to minimize a chip's exposure to radon throughout its fabrication, well before it is installed and operated. These decays may be a prominent source of ambient high-energy radiation as superconducting qubit platforms attempt to scale to larger-area devices, potentially mirroring the emergence of these decays as dominant backgrounds in large-scale rare-event searches~\cite{LinehanLZRadon2026,Aalbers2023a,Aprile2023,Collaboration2018f,ParveenThesis,Morrison2025RadonDaughterBackgroundsLZ}.

Because radon daughter plateout is a relatively slow process (100 - 500 atoms/cm$^2$/day~\cite{stein2018radon}), studying its impacts on superconducting qubit devices requires an acceleration of plateout rates on test devices. In this work, we present the design, operation, and testing of a facility providing such accelerated radon daughter plateout. In \S\ref{sec:RadonDaughterPlateoutBackgrounds}, we briefly introduce the $^{222}$Rn chain, the mechanics of daughter plateout on surfaces, and decays in that chain that are relevant to quantum computation. We follow with \S\ref{sec:RadonPlateoutApparatus}, in which we discuss the facilities for exposure and assaying radon-daughter activity, and \S\ref{sec:MeasurementsOfRadonDaughterPlateout}, in which we discuss a series of measurements characterizing the efficacy of the apparatus in boosting the rate of radon daughter plateout on devices and their packaging. We end with \S\ref{sec:projections}, in which we discuss projected implications of ambient radon plateout for quantum processors, before concluding in \S\ref{sec:conclusions}.

\section{$^{222}$Rn Daughter Plateout Process}
\label{sec:RadonDaughterPlateoutBackgrounds}

\begin{figure}[t!]
\centering
\includegraphics[width=1.0\linewidth]{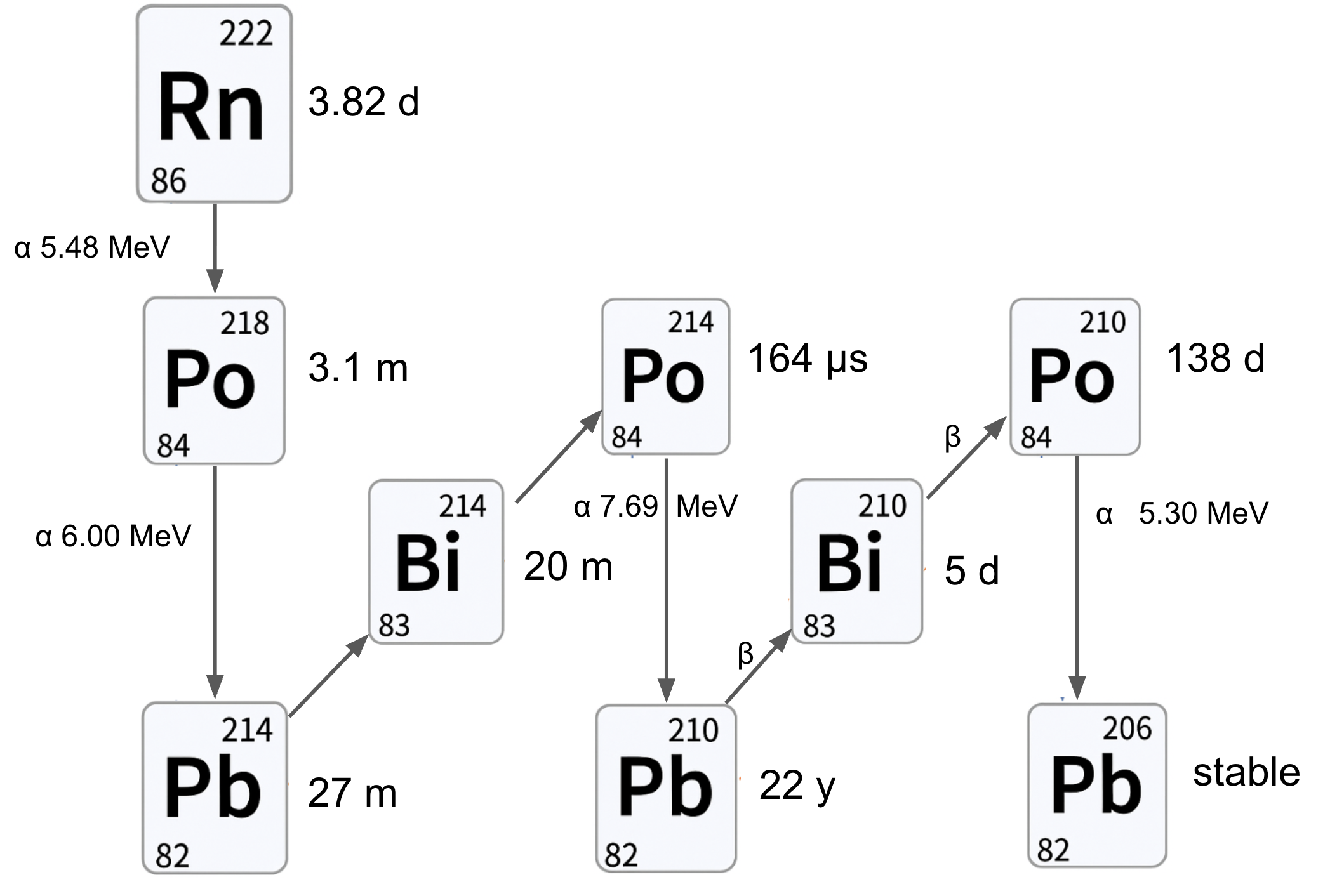}
\caption{$^{222}$Rn decay chain. Only prominent $\alpha$ decays (vertical arrows) are shown. $\beta$ decays are indicated by the upward-pointing diagonal arrows. }
\label{Figure: Radon_decay_chain}
\end{figure}

Gaseous $^{222}$Rn is produced in the decay chain of the naturally-occurring trace isotope $^{238}$U, which is widely prevalent in rocks and soil.  Owing to the relatively long, 3.8-day half-life of $^{222}$Rn, it diffuses into and accumulates in both indoor and outdoor environments~\cite{hopke1987radon,nazaroff1987radon}. In ambient air, typical concentrations give radon decay rates at the scale of 50~Bq/m$^{3}$~\cite{EPA_Radon_Action_Level}. However, depending on airflow and geological environment, the decay rate may be orders of magnitude larger~\cite{porstendorfer1994daily,kolarvz2009daily}. 

Upon decaying, $^{222}$Rn produces progeny (Fig.~\ref{Figure: Radon_decay_chain}) including $^{218}$Po and $^{214}$Po, which decay over $\mathscr{O}$(1 hour) to long-lived $^{210}$Pb ($T_{1/2}=22.2$~y). During this period, airborne progeny may drift to nearby surfaces and chemically attach, in a process called ``plate out.'' Due to the long half-life of the $^{210}$Pb left from $^{214}$Po decay, this plate-out process will yield a long-lived source of ionizing radiation local to the plate-out surface, which can be problematic if that surface belongs to a qubit device or the device-facing surfaces of its packaging.\footnote{$^{220}$Rn, a decay product of $^{232}$Th, is also naturally prevalent and unstable, but because all of its daughters are short-lived (half-lives up to 10 hours), any plate-out will not produce a long-term source of radiation.}

The rate and final locations of plate-out depend on the environment in which the decay sequence takes place. All radon progeny are subject to diffusion as they traverse the decay volume to a nearby surface, implying that the geometric size of the $^{222}$Rn-filled volume determines the time it takes for those daughters to reach surfaces, and as a result, the fraction of that progeny that arrives as $^{218}$Po, $^{214}$Pb, etc. Beyond simple diffusion, electric fields may also play a role: $\alpha$ decays of $^{222}$Rn, $^{218}$Po, and $^{214}$Po usually strip electrons from the recoiling nuclei~\cite{Knipp1941,wilkening1966radon, hopke1987radon, Hopke1996,LinehanLZRadon2026} while $\beta$ decays usually result in positively-charged progeny by the unity net increase in nuclear charge~\cite{wilkening1966radon}, giving these daughters a positive charge and making them subject to electrostatic drift in ambient electric fields. Other transport mechanisms, including convection~\cite{marlow1988electrical} and attachment to particulates in the air~\cite{Jacobi1972,Raabe1969,Porstendorfer1978,Otahal2025,Huet2001,Reineking1985,Guo2016} (which then themselves may be subject to electrostatic or convective transport), also impact plateout location.

Once a daughter reaches a surface, subsequent $\alpha$-decays prior to $^{210}$Pb (say, from plated-out~$^{218}$Po or $^{214}$Po) can cause their daughter nuclei to either ``embed'' $\mathscr{O}$(1-100 nm) into a surface~\cite{roos2002studies,morrison2018radon} or eject away from the surface. While embedded daughters may be impossible to remove via cleaning or wiping, they may be removed via chemical surface treatment~\cite{hoppe2007cleaning,zuzel2012removal2,Schnee2014a, guiseppe2018review,bruenner2021radon,zuzel2012removal}, while daughters on the surface (including those attached to particulates) may be removed via wiping~\cite{wojcik2007behavior, chernyak2025removal}. Each of these effects also displays a degree of surface material dependence~\cite{Chernyak2023,chernyak2025removal,stein2018radon}. Taken together, all of these considerations make modeling radon daughter plateout a highly complex problem that is difficult to accurately predict \textit{a priori}.

\begin{figure*}[t!]
\centering
\includegraphics[width=1.0\linewidth]{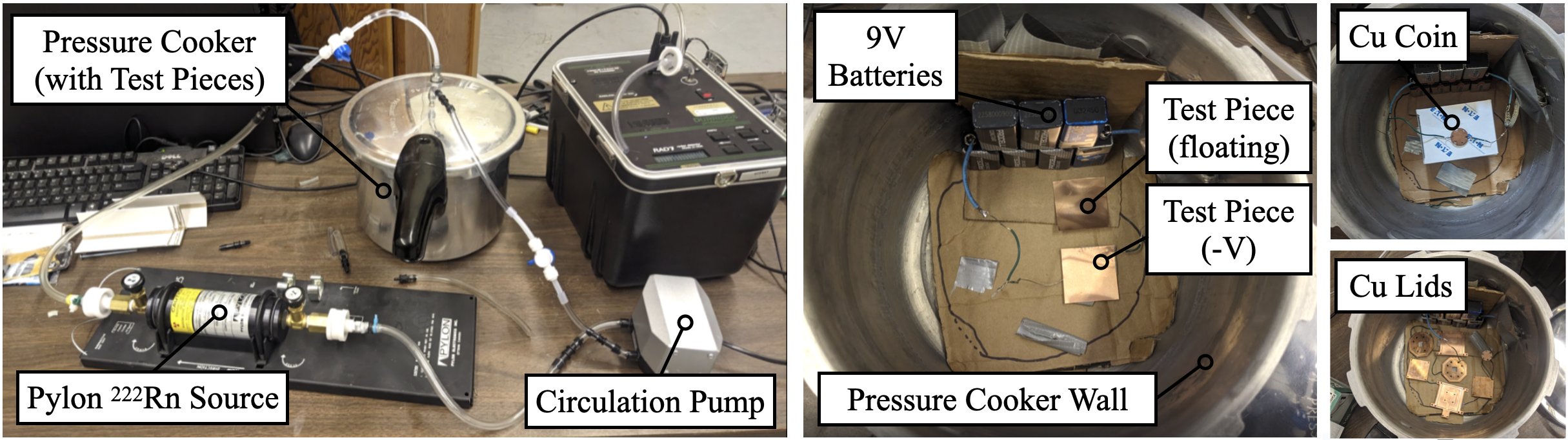}
\caption{Left: The radon exposure apparatus showing the flow-through $^{222}$Rn pylon source and circulation pump that drives the gas into the pressure vessel (exposure chamber). The radon concentration is measured with the \textsc{Rad7} radon detector. Right: Internal images of the pressure vessel showing exposure configurations for various work pieces. The cardboard provides insulation from chassis ground, enabling the at-field plate-out.}
\label{fig:PlateoutApparatusPlusLabeled}
\end{figure*}

The $^{210}$Pb that remains after all plateout and alpha decays from $^{218}$Po and $^{214}$Po have occurred, will remain for decades, creating energy depositions via its own decay ($Q=63.5$~keV) and those of its daughters $^{210}$Bi ($Q=1162.2$~keV) and $^{210}$Po ($Q=5407$~keV). Since the $^{210}$Pb half-life (22.2~y) is much longer than those of $^{210}$Bi (5~d) and $^{210}$Po (138~d), the decay rates of the two daughters will increase until they reach secular equilibrium with $^{210}$Pb over several half-lives of $^{210}$Po, subsequently tracking downward with the same decay timescale of $^{210}$Pb.  This ``grow in" process takes approximately 1 year, and leaves a quasi-permanent source of $\beta$ and $\alpha$ decays that will be present over the entire lifetime of a device and which may contribute to the ionizing radiation background capable of creating correlated errors in qubit devices.

In this work, we focus primarily on understanding surface activities of $^{210}$Po, as that isotope's high-energy alpha decay is likely most disruptive to coherence of qubit systems. Given the constantly varying rate of these decays, it is useful to define a metric for this late-time $^{210}$Po surface activity as its activity three years after $^{222}$Rn exposure, which we call $\mathcal{A}_{3y}$. This metric has the benefit that enough time has passed to clearly establish secular equilibrium, while still reflecting reasonable operating timescales of QIS devices.

As a point of comparison to the experimental measurements presented later in this work, we estimate the rate of ambient $^{222}$Rn progeny plateout on surfaces (as well as the resulting $\mathcal{A}_{3y}$) using a ``dead-air'' model (Appendix~\ref{appendix:deadair}). In this model there is no mechanism for removal of radon daughters from the gas other than plate-out (\textit{i.e.}, no ventilation). We define ambient conditions as a steady-state radon concentration of 50 Bq/m$^3$, for which our model estimates $\approx110$ deposited $^{210}$Pb atoms per cm$^2$ per day, which yields $\mathcal{A}_\mathrm{3y}^\mathrm{model}\approx0.04$ mBq/cm$^2$. This model is roughly consistent with measurements of this rate at SNOLAB, an underground laboratory where the radon concentration is about 135 Bq/m$^3$, which yielded average $^{210}$Pb deposition rates of 249 and 423 atoms/cm$^2$/day on polyethylene and copper, respectively~\cite{stein2018radon}.

\section{Radon Plateout and Measurement Facilities}
\label{sec:RadonPlateoutApparatus}

For this work, we use an enhanced radon exposure system (Fig.~\ref{fig:PlateoutApparatusPlusLabeled}) to accelerate radon plateout on test pieces. The system is comprised of three components: a Pylon RN-1025 $^{222}$Rn source, a dedicated circulation pump, and a 3.5 liter pressure vessel whose walls are impervious to Rn diffusion. These elements are connected in a closed loop via nylon tubing, which allows for injection of $^{222}$Rn into the sealed pressure vessel, within which test samples are seated for $^{222}$Rn daughter plateout.

A typical exposure cycle consists of a transient injection of fresh $^{222}$Rn followed by a long, no-flow plateout period during which $^{222}$Rn daughters plate out onto the test pieces and interior surfaces of the vessel. During the injection cycle, the flow rate in the closed loop is set to 2 liters of gas per minute and the injection is carried out for 2 minutes so as to provide reasonably uniform mixing within the pressure vessel. After this period of flow, the pump is stopped and the pylon source sealed. The injections are only performed once the pylon has been isolated for at least seven days. This timescale, corresponding to a few $^{222}$Rn half-lives, is also the period over which the samples are exposed to the radon-laden air in the pressure vessel.

At the end of an injection period, the achieved $^{222}$Rn concentration in the vessel is between 4 and 7 MBq/m$^3$, as measured by a \textsc{Rad7} radon detector~\cite{DurridgeRAD7Manual}, which infers $^{222}$Rn activities from the equilibrium activity of $^{218}$Po over a sampling period of about 30 minutes. While the \textsc{Rad7} systematically under-measures the true rate above volume activities of 750 kBq/m$^3$ due to event pileup, we correct for this effect using a set of auxiliary measurements (see Appendix~\ref{app:rad7-saturation}). The average achieved 8.9 MBq/m$^3$ (corrected) $^{222}$Rn concentration is about 178,000$\times$ larger than typical indoor, ambient-air concentrations~\cite{EPA_Radon_Action_Level}.

Once a test piece has been exposed to radon daughter plateout, a separate $\alpha$-detector facility is used to measure the level of radon daughter contamination via measurements of the $^{210}$Po decay. The $\alpha$ detector used in this work is an Alpha Duo (Fig.~\ref{fig:alphacounter})~\cite{OrtecAlphaDuoManual}, a dual-chamber spectrometer with pressure-regulated vacuum chambers and 1.69-cm-radius silicon detectors for $\alpha$ detection. Each chamber's detector has an energy calibration that is good to 1\% (i.e. $\sim$50 keV) at the $^{210}$Po alpha energy and a background rate of approximately 0.25 counts per hour per 10 keV. 

\begin{figure}[t!]
    \centering
    \includegraphics[width=0.995\linewidth]{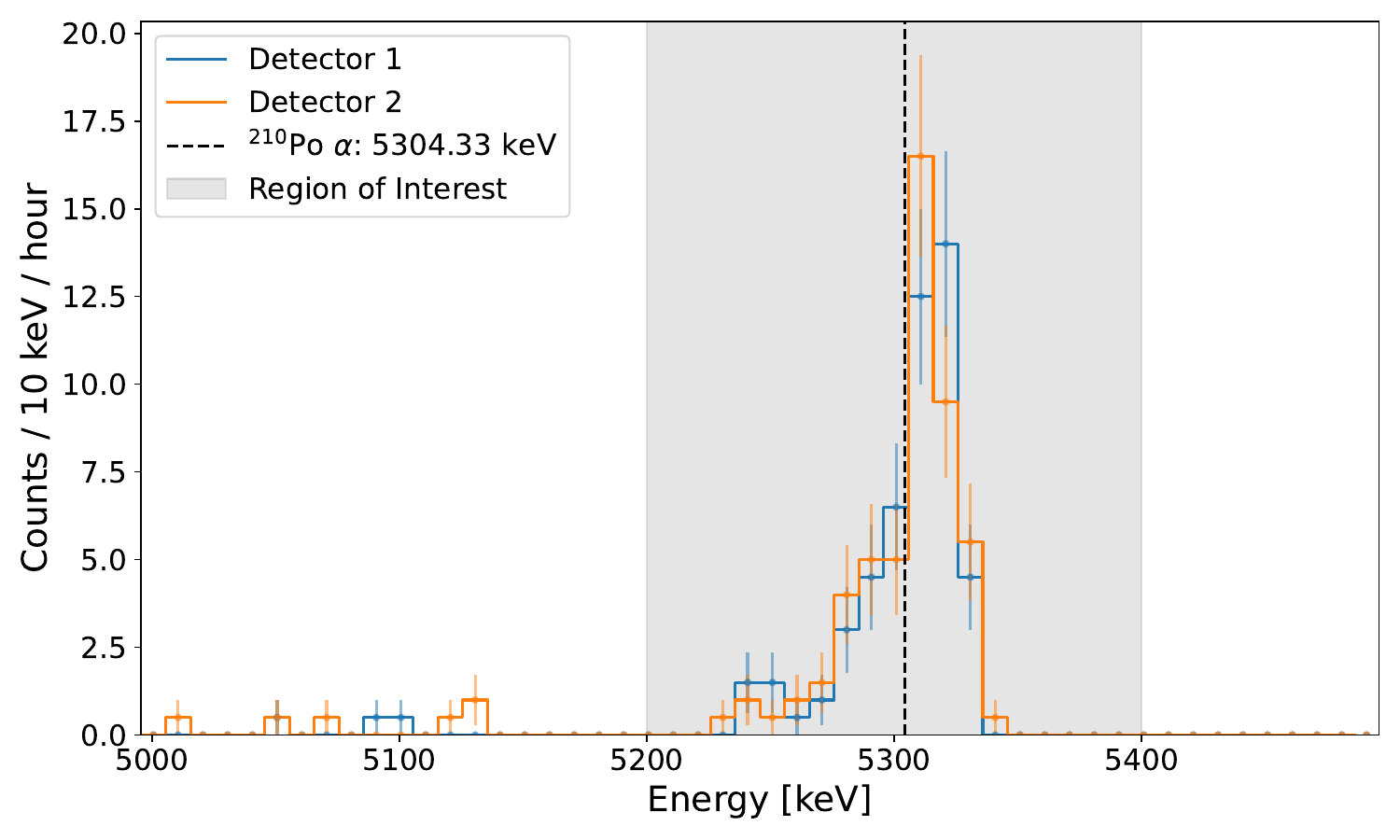}
    \caption{$^{210}$Po alpha spectrum for the same copper sample measured in the two detectors of the \text{Alpha Duo}. In both cases, the top surface of the strip was 4 mm away from the detectors. These data were acquired two hours apart, so the grow-in of $^{210}$Po during measurements is minimal.} 
    \label{fig:Cu-strip-measurements-across-different-detectors} 
\end{figure}

A measurement cycle comprises loading test pieces into the Alpha Duo, evacuating the sample spaces to $<2$ torr, and taking data. For each measurement, a spectrum like that in Fig.~\ref{fig:Cu-strip-measurements-across-different-detectors} is obtained, with a peak at the 5.3 MeV $^{210}$Po endpoint accompanied by a tail to lower energy. This tail is a product of two effects linked the variation in alpha decay direction. First, varying alpha incidence angles on the Alpha Duo cause variations in energy lost in the detector's $\sim$50~nm-thick dead layer. Second, variations in emission angle from embedded daughters induce variations in alpha energy lost in the sample itself. A simple rate for any given measurement is extracted by integrating the spectrum between [5200, 5400] keV. Detection efficiencies vary between samples due to variation in sample geometry and the solid angle of the detector a sample subtends (Appendix~\ref{appendix:alphacounter}). These efficiencies are divided out of all measurements in this text to estimate the total plateout activity on each piece.

\section{Measurements of Radon Daughter Plateout}
\label{sec:MeasurementsOfRadonDaughterPlateout}

To assess the efficacy of this plateout facility, we exposed and subsequently measured $^{210}$Po activities of several copper test pieces, the properties of which are summarized in Table~\ref{table:copper-samples}. Each test piece was measured with the goal of probing plateout rates' dependence on various system conditions, as discussed below. In all cases, we compare results to the dead-air model (Appendix~\ref{appendix:deadair}) assuming an ambient $^{222}$Rn activity of 50~Bq/m$^{3}$, representative of the piece existing under laboratory benchtop conditions in air that has not been enriched with (or reduced of) $^{222}$Rn. We use $\mathcal{A}_{3y}$ as the primary metric of comparison for these studies.

\begin{table}[b]
 \small
 \centering
 \begin{tabular}{|c|c|c|}
 \hline
 \textbf{Sample} & \textbf{Exposure Bias} & \textbf{Dimensions} \\
 \hline
 \hline
 Square Cu Foil ``A" & No applied field & $4\times4$ cm$^2$ \\
 Square Cu Foil ``B" & No applied field & $4\times4$ cm$^2$ \\
 Square Cu Foil ``C" & Floating         & $4\times4$ cm$^2$ \\
 Square Cu Foil ``D" & -63~V            & $4\times4$ cm$^2$ \\
 Cu ``Coin"              & -63~V            & $r = 0.95$ cm     \\
 \hline
 \end{tabular}
 \caption{Summary of the copper samples under test in \S\ref{sec:MeasurementsOfRadonDaughterPlateout}. The dimensions are also provided for converting areal surface activity density to total surface activity.}
\label{table:copper-samples}
\end{table}

\subsection{Copper Foil Tests: No Field}
\label{subsec:ThinFoilTestsNoField}

\begin{figure}[t]
\includegraphics[width=1.0\linewidth]{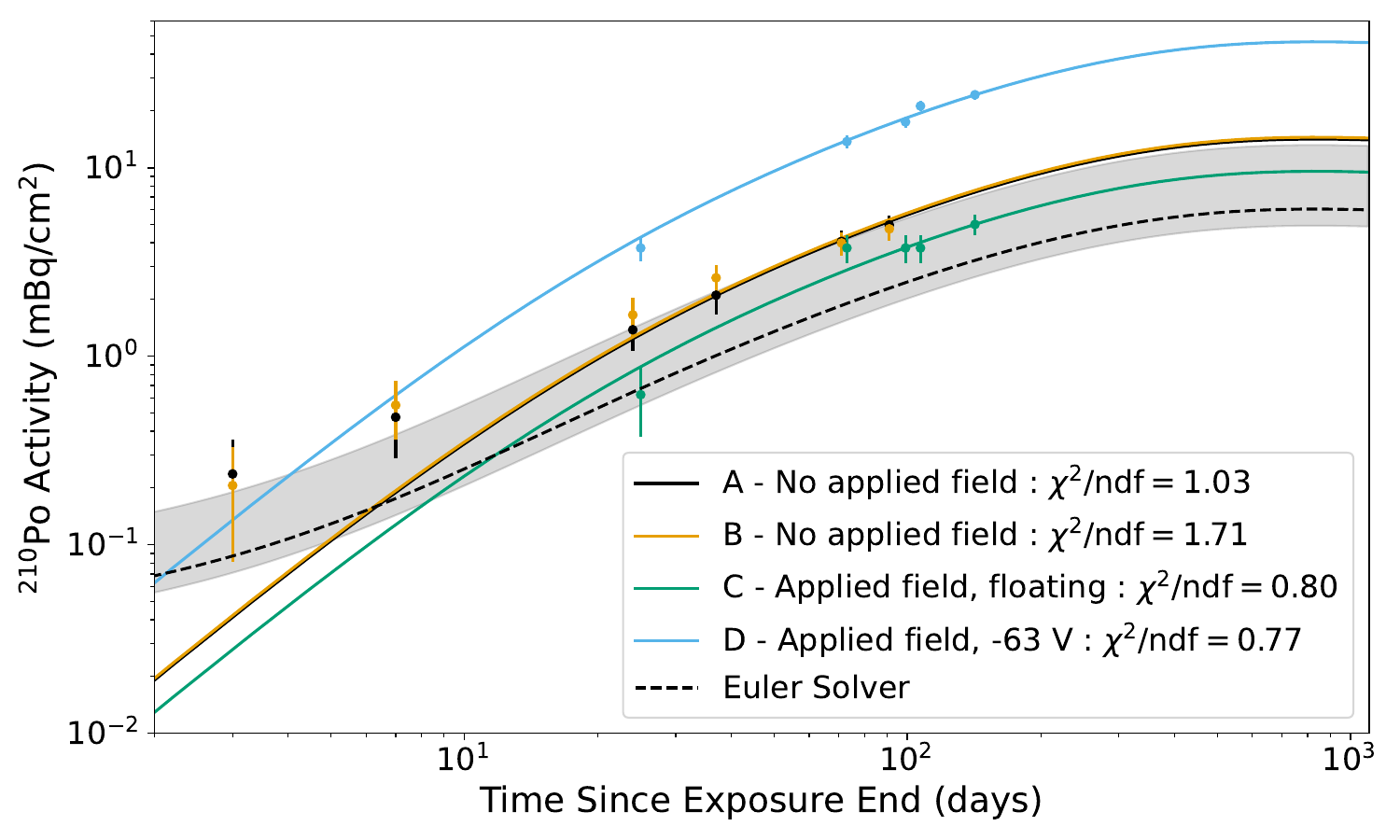} 
\caption{Measurements of total surface activity (per sq. cm) of Cu Foils ``A", ``B", ``C", and ``D". Each pair (A \& B and C \& D) was placed in the radon exposure vessel separately for 7 days. All solid curves are best fits to the simplified Bateman Equation (Eq.~\ref{eq:finalBatemaneq}), in which the only free parameter is the initial activity of $^{210}$Pb, and the initial surface activities of $^{210}$Bi and $^{210}$Po are fixed to be zero. Also shown is a forward-Euler solution to Eqs.~\ref{eq:ODE-gas-base} and~\ref{eq:ODE-surf-base} (black dashed), modified to allow for a decaying $^{222}$Rn concentration, for a 7-day exposure (with $\rho_A=7.3$ MBq/m$^3$ and $A/V=38.2$ m$^{-1}$) which qualitatively captures the early time behavior (seen in A and B), whereas the simplified grow-in model (Eq.~\ref{eq:finalBatemaneq}) underestimates that period but accurately captures $\mathcal{A}_\mathrm{3y}$. This Euler-solver was not tuned to match the data, and the surrounding gray band indicates the range of corrected initial activities.} 
\label{fig:Cu-strip-combined}
\end{figure}

We first perform a test of the baseline performance of this facility by exposing a pair of (4$\times$4)-cm$^2$ thin copper foils, named Foil~A and Foil~B, which were left in contact with the electric ground of the exposure vessel. These were exposed to radon for 7 days, with an initial inferred radon concentration of 7.3~MBq/m$^{3}$. After exposure, the foils were measured over several weeks with the alpha counter, and yielded $^{210}$Po rates as shown in Fig.~\ref{fig:Cu-strip-combined}. The expected grow-in (solid lines in Fig.~\ref{fig:Cu-strip-combined}) is evident, and a fit to a Bateman model (Eq.~\ref{eq:finalBatemaneq}) yields an $\mathcal{A}_{3y}\approx13$~mBq/cm$^{2}$, a factor of $1.8\times10^4$ larger than plateout from ambient conditions as predicted by the dead-air model, and in good agreement with the dead-air model prediction for these conditions ($\mathcal{A}_\mathrm{3y}^\mathrm{model}=10.6$ mBq/cm$^2$). 

\subsection{Thin Foil Tests: Plateout E-Field Dependence}
\label{subsec:ThinFoilTestsEField}

An appreicable fraction of $^{222}$Rn daughters are expected to be positively charged by the chain of $\alpha$ and $\beta$ decays.  Hence, ambient electric fields (e.g. as may arise with static electricity) are expected to impact the probability of daughter isotope plateout on surfaces, especially those that are deliberately held at voltage~\cite{Chernyak2023}. In this way, by applying a negative bias voltage with respect to the grounded chassis to our samples, we may further accelerate plate-out by electrostatically attracting daughters to them.

To isolate this E-field dependence, we placed two new 4-cm-square copper foils, named Foil~C and Foil~D, identical in form to those described in Section~\ref{subsec:ThinFoilTestsNoField}, on a layer of insulating cardboard in the pressure vessel (Fig~\ref{fig:PlateoutApparatusPlusLabeled}, center). We set Foil~D at -63~V with respect to ground at the vessel wall, and left the Foil~C floating. We exposed this configuration to air with an initial $^{222}$Rn concentration of 7.3~MBq/m$^{3}$ for 7 days. Resulting plate-out alpha rates are shown in Figure~\ref{fig:Cu-strip-combined}, and indicate $\mathcal{A}_{3y}=50$~mBq/cm$^2$ for the foil held at negative voltage, an enhancement factor of $\approx$5 over the foil left electrically floating. While this enhancement factor qualitatively agrees with the trend observed in Ref.~\cite{Chernyak2023}, the effect of a field will also depend on the geometry of the exposure vessel and test pieces, the voltages applied, and the flow patterns involved in the exposure environment~\cite{LinehanLZRadon2026}. With this electric-field acceleration, our apparatus reaches a plate-out rate enhancement of $7\times10^4$ relative to predictions of the dead-air model under ambient conditions for the same exposure period (7 days).

We also note that the ``floating" foil, ``C", shows a deficit in activity compared to the no-field foils (``A" \& ``B"). This is expected as the field preferentially directs Rn-progeny ions in the gas toward the biased foil (``D"), and away from the floating foil.

\begin{figure}[t]
\centering
\includegraphics[width=\linewidth]{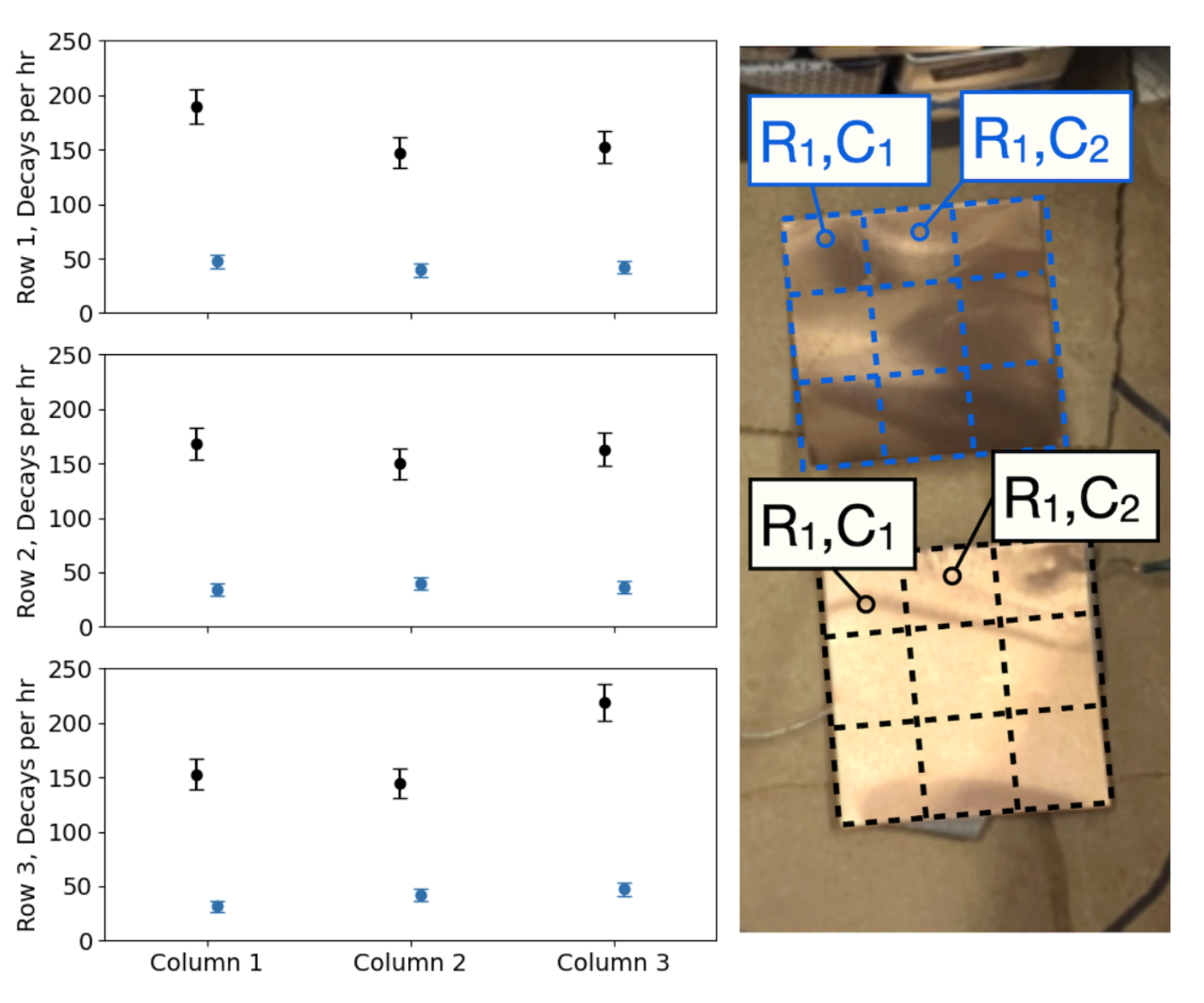}
\caption{Left: Rates of the nine sub-squares of Foil~C (floating, blue) and Foil~D (-63~V, black), demonstrating uniformity of plateout across both. Right: Diagram showing orientation and clocking of the rows and columns referenced in the plots on the left. }
\label{fig:SpatialUniformityResults}
\end{figure}

We subsequently probed the spatial uniformity of plateout on Foils C and D. Each foil was cut into nine identical 1.33~cm x 1.33~cm squares, and for each square the $^{210}$Po activity was measured. The results are shown in Fig. \ref{fig:SpatialUniformityResults}. Here, these activities are the activities at the time of measurement (i.e., not $\mathcal{A}_{3y}$), and demonstrate that plate-out is nearly uniform within statistical errors, even for pieces held at voltage.

\subsection{Tests of Surface Cleaning}
\label{subsec:SurfaceWiping}

While some radon progeny may be found ``embedded'' in a surface from \(^{218}\)Po and \(^{214}\)Po decays, a nonzero fraction of plated-out \(^{210}\)Pb-chain daughters will be only weakly bound to the surface and therefore removable with wiping or rinsing~\cite{Schnee2014a,Zuzel2018}. This removable progeny may take the form of individual daughters bound to a surface through weak Van der Waals forces~\cite{Leung2005}, or in some environments may also take the form of daughters attached to ambient particulates which have landed on a surface~\cite{Jacobi1972,knutson1983radon,vanmarcke1985equilibrium,national1991comparative}. To better predict the long-term activity level of our exposed samples, we devised tests to probe the removable fraction of daughters resulting from the exposures. 

These tests involved studying the effect of successive wipes on a copper qubit housing lid, named the ``coin" based on its size and shape, as shown in Fig.~\ref{fig:PlateoutApparatusPlusLabeled}. The copper coin was exposed in our setup with a -63~V bias applied. The coin was exposed prior to the foil studies and at a time when we had less consistent exposure procedures.   Following its somewhat uncertain exposure history to make projections of $\mathcal{A}_\mathrm{3y}$ is challenging.  However, it is also not necessary for studying the removal of Rn progeny so we simply measured its activity level immediately following exposure.  At the start of the study, the coin was measured to have a $^{210}$Po activity of 293 $\pm$ 9 mBq/cm$^2$, prior to any wiping.

\begin{figure}[t]
\centering
\includegraphics[width=\linewidth]{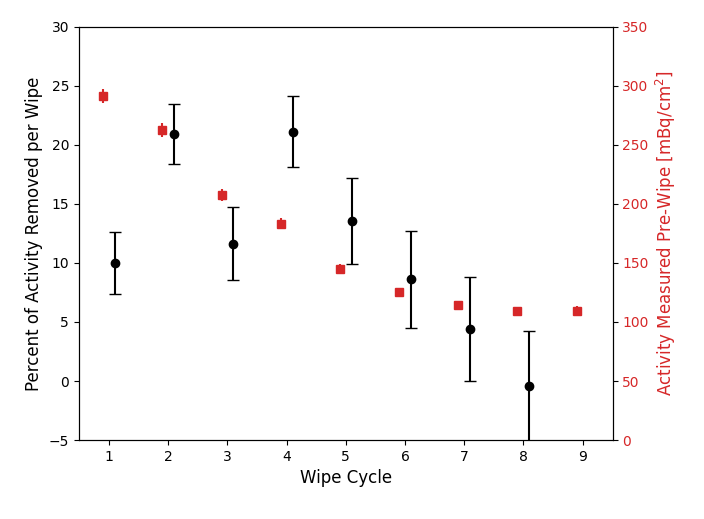}
\caption{Absolute $^{210}$Po activity of the copper coin before each wipe, and relative removal per wipe trial, shown slightly offset from the wipe number abscissa. The first trial was done with a dry wipe. The remaining trials used alcohol-soaked wipes. Uncertainty bands for the pre-wipe activity measurement (red) are shown, and are comparable to the size of the marker. An artificial Wipe Cycle 9 is included to show the absolute activity \textit{after} Wipe Cycle 8 (the final wipe cycle).}
\label{fig:RemovalPerWipe}
\end{figure}

The wiping procedure is as follows. The first wipe cycle involved only a single dry Kimtech wipe. On the same day, the second wipe cycle involved five wipes with an isopropyl-alcohol-saturated Kimtech wipe. After five days, a set of additional wipe cycles were performed, each of which was performed approximately one day after the preceding wipe cycle. Each of the following wipe cycles involved wiping the surface 5 times with a Kimtech wipe saturated with isopropyl alcohol. After each wipe cycle, the piece was re-measured in the alpha counter. While a small amount of \(^{210}\)Po grow-in occurred between successive wipe cycles, the short, few-day inter-wipe period limited this grow-in to less than 5\(\%\) prior to each subsequent wipe. Results are shown in Fig.~\ref{fig:RemovalPerWipe}. After wipe cycle 6, the coin was dipped in isopropyl alcohol to see whether the remaining removable activity could be rinsed away. After this treatment, the fraction of \(^{210}\)Po removed by subsequent wipes was measured to be statistically consistent with zero at the \(1\sigma\) level. We compute the fraction of the initial \(^{210}\)Po remaining after all wipes as the final absolute activity divided by the initial absolute activity, and find this to be \(0.38\pm0.06\). Here we quote a systematic uncertainty reflective of the grow-in over the full eleven-day wipe process, which is likely to cause our measured value to overestimate the fraction remaining. This suggests a removable (surface) fraction of approximately 62\% for plated-out daughters from our exposure facility, modestly higher than that found in Ref.~\cite{chernyak2025removal} using acetone.

While the precision of the measured 62$\%$ removable fraction is likely to be somewhat dependent on detailed geometry of the samples and other conditions in the plateout apparatus, these results nevertheless grant an estimate of the on-surface fraction of daughters which can be used to inform models of the radiation response of superconducting qubits and other devices with fragile surface features (like Josephson junctions) which cannot be safely mechanically wiped. However, if one can rinse a device or its packaging with a solvent such as isopropyl alcohol without degrading its subsequent performance, radon backgrounds may be suppressed by a factor of two or more~\cite{Zuzel2018}. More aggressive surface cleaning strategies such as electropolishing or etching have been shown to remove even the embedded daughters not removable with mechanical wiping~\cite{Schnee2014a,Guiseppe2018}. While unrealistic to perform on finished devices, these strategies may be readily applied to device enclosures, and may also be used on raw device materials prior to fabrication to reduce the integrated radon progeny buildup in a final device.

\section{Projected Effects of Radon at Ambient Concentration}
\label{sec:projections}

The success of our dead-air model in predicting rates in \S\ref{sec:MeasurementsOfRadonDaughterPlateout} motivates us to estimate the impact of radon plateout on the rate of ionizing events in qubit devices under ambient conditions. For these projections, we define a benchmark device geometry, consisting of a $5\times5\times0.5$\,mm$^{3}$ silicon substrate (representing the qubit chip) and a copper enclosure surrounding this.  We assume a uniform activity density for $^{210}$Po at 0.04 mBq/cm$^2$.  This is informed by the dead-air model predicted value of $\mathcal{A}_\mathrm{3y}$ assuming a ``typical room-air" ambient Rn concentration of 50~Bq/m\(^{3}\). The calculation assumes that this device is exposed for one year before being installed in a vacuum environment typically used in qubit operation. We consider two potential effects: transient energy depositions from all \(^{210}\)Pb-chain decays, and crystal defect production from the set of ion recoils occurring throughout the \(^{222}\)Rn chain.

\subsection{Transient Energy Depositions}
\label{subsec:TransientEnergyDepositions}

\begin{figure}[t]
\centering
\includegraphics[width=\linewidth]{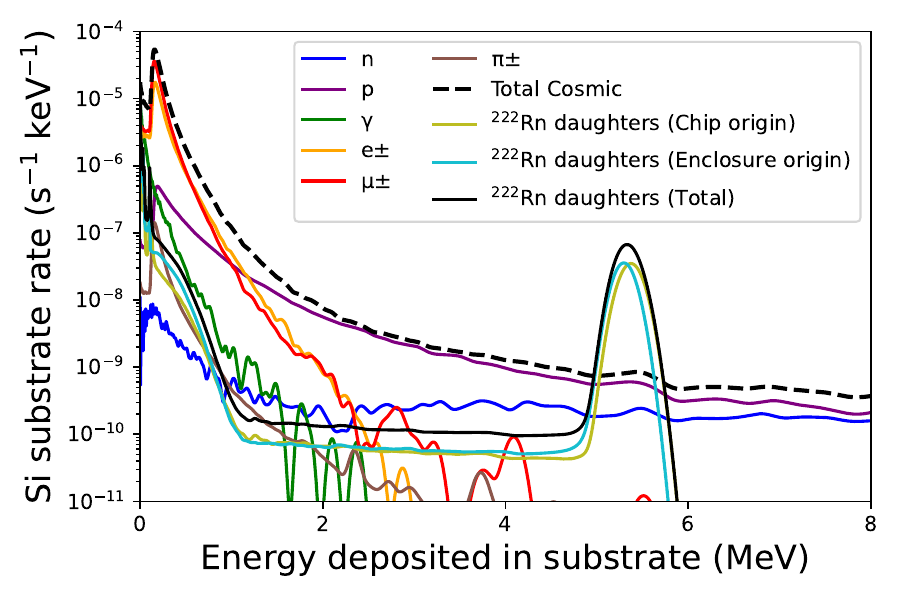}
\caption{The simulated reconstructed energy spectrum of $^{222}$Rn progeny decays in a $5\times5\times0.5$ mm$^{3}$ silicon chip, originating from the chip surfaces (chartreuse) and the enclosure surfaces (cyan). The total spectrum from these decays is shown by the solid black line. This rate is based on an assumption of $^{210}$Po surface activity density of 0.04 mBq/cm$^2$ for both the chip surfaces and enclosure surfaces. Also shown are the spectra for cosmic rays (reproduced from Ref.~\cite{Fowler2024}) for comparison. We apply to the $^{222}$Rn progeny spectra a smearing function identical to that used in Ref.~\cite{Fowler2024}.}
\label{fig:G4-spectra}
\end{figure}

Decays within the long-lived \(^{210}\)Pb-chain will lead to ionization, phonon production, and transient quasiparticle poisoning events within superconducting qubit devices. To estimate the energies deposited by this chain in our benchmark device geometry, we simulate \(^{210}\)Pb-chain decays from these pieces using \textsc{Geant4}, a Monte Carlo particle simulation package~\cite{Allison2016,Allison2006,Agostinelli2003}, and then we reconstruct the total energy deposited per-event in the silicon substrate (Appendix~\ref{appendix:g4-sims}). Figure~\ref{fig:G4-spectra} shows the energy spectrum of these decays in comparison to contributions from cosmic rays~\cite{Fowler2024} for a laboratory on the surface of the Earth. We separately show contributions from plate-out on the chip surfaces and from plate-out on the enclosure surfaces, because in practice these can have different radon exposure histories. Each of these contributions is a sum over decays from $^{210}$Pb, $^{210} $Bi, and $^{210}  $Po assumed to be in secular equilibrium. The excess at low energies, below $\approx$1~MeV, is a result of $\beta$'s and $\gamma$'s from the decays of $^{210}$Pb and $^{210}$Bi, together with the $^{206}$Pb ion recoils from $^{210}$Po decay. The peak at 5.3~MeV is from the $^{210}$Po $\alpha$-decay, with contributions from fully-contained decays in which the entire $Q=5.4$ MeV is deposited in the substrate.

We also present, in Table~\ref{tab:integrated-event-rates}, the integrated event rates, from Fig.~\ref{fig:G4-spectra}, with deposited energy $>1$ MeV, which we take to be the order-of-magnitude energy scale above which gap engineering cannot efficiently prevent $T_1$ errors~\cite{Pinckney2026}. For a chip operated in a surface laboratory with a year of ambient Rn exposure, the rate of events from $^{222}$Rn progeny is nearly equal to the cosmic-ray proton flux.  The latter is hypothesized to be a major contributor to the low-rate population of correlated-errors that persists in gap-engineered qubit systems~\cite{Acharya2024,Pinckney2026}. While the relative contributions of these cosmic ray components and radon daughter plate-out depend on a given device's exposure history and may be smaller or larger than those presented in Fig.~\ref{fig:G4-spectra}, this nonetheless demonstrates that without careful exposure limitations for a device and its packaging, radon plate-out may be a non-negligible component of high-energy impacts in a qubit device operated in a surface lab.

If one attempts to mitigate cosmic ray backgrounds by operating underground, the importance of these radon daughter plate-out backgrounds becomes even more stark. For example, at a depth of 225 m.w.e characteristic of a shallow underground lab~\cite{Michael2008}, cosmic ray protons and neutrons produced in hadronic showers in the atmosphere are effectively eliminated~\cite{AalsethShallowSite, PDG2024CosmicRays, ChatterjeeNeutrons}, and the remaining muon flux is reduced by $\approx99.5\%$~\cite{Mei2006}\footnote{The remaining comsic-ray components (e$^\pm$, $\gamma$, $\pi^\pm$) are also entirely attenuated by a shallow rock overburden~\cite{Beatty2020PDGCosmicRays}, though interactions of muons in the rock can produce additional flux of electrons, photons, and hadrons.}. This will make the alpha peak in Fig.~\ref{fig:G4-spectra} dominant even for substantially lower exposure times than what we assume. This tendency of radon backgrounds to increase in relevance with additional shielding techniques and devices at scale mirrors observations from the dark matter direct detection and other rare event search communities~\cite{LinehanLZRadon2026,CRESST,CRESSTRoughness,CUORE-0}. We note that other high-energy backgrounds, including those from intrinsic \(\alpha\)-decaying impurities in chip packaging and neutrons from cosmic ray muon spallation, may also be present in this energy range and relevant for underground operation. They will have either location- or material- specific rates, and we do not attempt to quantitatively model these here.

\begin{table}[t]
\begin{tabular}{|l|c|c|}
 \hline
 & \multicolumn{2}{c|}{\textbf{Integrated Rate }[s$^{-1}$]} \\
 \cline{2-3}
 \textbf{Source}& \textbf{$E_\mathrm{dep}>1$ MeV}& \textbf{$E_\mathrm{dep}>2.6$ MeV} \\
 \hline
 \hline
  CR p & $2.1 \times 10^{-5}$ & $4.6 \times 10^{-6}$ \\
  CR e$^\pm$ & $1.1 \times 10^{-5}$ & $9.6 \times 10^{-9}$ \\
  CR $\mu^\pm$ & $7.2 \times 10^{-6}$ & $1.3 \times 10^{-7}$ \\
  CR n & $2.5 \times 10^{-6}$ & $2.0 \times 10^{-6}$ \\
  CR $\gamma$ & $5.1 \times 10^{-7}$ & 0 \\
  CR $\pi^\pm$ & $2.1 \times 10^{-7}$ & $3.7 \times 10^{-8}$ \\
  \hline
  $^{222}$Rn prog. & $2.2 \times 10^{-5}$  & $2.1 \times 10^{-5}$ \\
  \hline
  \end{tabular}
  \caption{Integrated event rates (above 1 MeV and 2.6 MeV) in a surface facility for cosmic ray (CR) components~\cite{Fowler2024} compared to that of $^{222}$Rn progeny (specifically, $^{210}$Pb, $^{210}$Bi, $^{210}$Po, though nearly all the activity in this energy range is contributed by the latter). 1 MeV was chosen as an order-of-magnitude energy scale above which gap engineering begins to wane in efficacy.  2.6 MeV is the highest energy $\gamma$-ray emitted by the most common sources of environmental radiation. As $\gamma$-ray's are highly penetrating, these particles are typically the dominant component of environmental radiation incident on a device.  } 
  \label{tab:integrated-event-rates}
\end{table}

\subsection{Crystal Defect Production}
\label{subsec:CrystalDefectProduction}

Generally, uncontrolled damage to a device's crystalline structures should be avoided, as it can degrade qubit performance. For example, a limitation of superconducting qubits is the presence of quantum two-level systems (TLS) with energy splittings near typical \(\mathcal{O}(10\mu\)eV) qubit transition frequencies~\cite{MullerTLSReview,ShaliboTLS,Muller2015Fluctuations,KlimovFluctuations,Lisenfeld_2016,qTLS,Thorbeck:2022yzs}. These often arise in oxide layers (including Josephson junction barriers) and other regions where atomic-scale disorder is present~\cite{MullerTLSReview}, and can stochastically modify a qubit's coherence times and transition frequency.  Studies indicate that damage induced by ion bombardment, including structural modifications of oxide layers~\cite{ArgonMillingVanDamme} and crystal amorphization at an interfacial substrate surface~\cite{QuintanaFabTLS} may increase TLS noise.

Rn daughters implanted in solid-state devices can also give rise to crystalline defects, known as Frenkel defects. They are created when a heavy, recoiling daughter ion knocks an atomic nucleus out of the crystal lattice into a space between other lattice atoms.  The subsequent displaced ion and hole form a pair known as a Frenkel defect~\cite{Frenkel}. These defects can arise either in the ``embedding'' process undergone by \(^{214}\)Pb and \(^{210}\)Pb ions during decays of \(^{218}\)Po and \(^{214}\)Po in ambient air, or from ejection of \(^{206}\)Pb ions during \(^{210}\)Po decay after plate-out. Both the alpha and recoiling ion create a track of displaced ions and vacancies, and can give sufficient energy to the displaced ions that they create their own defects, thus forming a cascade. 

To study this effect, we used the \textsc{SRIM} simulation package~\cite{Ziegler:SRIM} to impinge $^4$He ($\alpha$), $^{214}$Pb, $^{210}$Pb, and $^{206}$Pb ions onto various materials.  We recorded the average range and average number of vacancies produced by each ion along its track. The results show that while the \(\alpha\)'s tend to have ranges much longer than the thickness of an \(\mathcal{O}(100~\textrm{nm})\)-thick film, the heavy ion recoils are fully contained, creating a denser set of vacancies in a local thin-film area. 
Table~\ref{tab:srim-results} shows example data for the properties of $^{206}$Pb tracks in materials of relevance to superconducting qubit devices.  Using the vacancy-per-ion multiplicities generated in the simulation, we calculate a total number of vacancies produced by \(^{218}\)Po and \(^{214}\)Po decays assuming an exposure of one year at 50~Bq/m\(^{3}\) ambient radon volume activity\footnote{Here, to provide an upper limit, we assume that 100\(\%\) of \(^{218}\)Po is deposited on a surface before it decays, though the real fraction is expected to be lower.}, as well as the total number of vacancies produced per year from \(^{210}\)Po decays during device operation. This yields \(\mathcal{O}(10^{4})\) tracks creating up to \(\mathcal{O}(10^{7})\) defects.

 While this number is large, it is still many orders of magnitude below the number created in the ion bombardment studies discussed above (with typical incident ion fluxes of $10^{13}$--$10^{17}$ cm$^{-2}$ s$^{-1}$ up to 10 keV).  Thus it is unlikely that ambient radon plateout will create broad, uniform regions of damage in crystal structures. However, for sufficiently energetic and heavy (relative to the substrate components) impinging ions, it is possible for long chains of defects to create local pockets of disorder and amorphization~\cite{PelazAmorphization,CamaraAmorphization}.  This may yet be problematic for future qubits, once the more prominent sources of TLS noise are resolved.

\begin{table}[t]
    \renewcommand{\arraystretch}{1.25}
    \begin{tabular}{|lrrrrrr|}
    \hline
     & \multicolumn{6}{|c|}{\textbf{\(^{210}\)Po decays}}\\
    \hline
    \hline
    \multicolumn{1}{|l|}{Particle} &
    \multicolumn{3}{c|}{\textbf{$\alpha$} (5304~keV)} &
    \multicolumn{3}{c|}{$^{206}$Pb (103~keV)}\\
    \cline{1-7}
    \multicolumn{1}{|l|}{Target} &
    \multicolumn{1}{c}{Si} &
    \multicolumn{1}{c}{Nb} &
    \multicolumn{1}{c|}{Al} &
    \multicolumn{1}{c}{Si} &
    \multicolumn{1}{c}{Nb} &
    \multicolumn{1}{c|}{Al} \\
    \multicolumn{1}{|l|}{Range {[}$\mu$m{]}} &
    26.6 & 11.5 & \multicolumn{1}{c|}{23.4} &
    0.047 & 0.018 & \multicolumn{1}{c|}{0.040} \\
    \multicolumn{1}{|l|}{Vacancies/ion} &
    339 & 0.3 & \multicolumn{1}{c|}{0.1} &
    2314 & 2042 & \multicolumn{1}{c|}{1518} \\
    \hline
    \end{tabular}
    \caption{Results of the SRIM simulations for $^{210}$Po decays. The targets were simulated with thickness 100 $\mu$m for substrates (Si) and 100 nm for superconductors (Nb, Al). 1,000 ions were simulated for each target/ion pair.}
    \label{tab:srim-results}
\end{table}

\section{Conclusions}\label{sec:conclusions}

Plate-out of $^{222}$Rn daughters, long identified as a significant background for dark matter direct detection and rare event searches, is a source of high energy, \(>\)MeV-level particle impacts that scale with device area and will lead to charge, phonon, and quasiparticle bursts in superconducting qubits.  To date, no studies have focused on this background as a dominant source of correlated error bursts in superconducting processors.  However, this paper demonstrates that with reasonable assumptions about a device handling history, it is possible for Rn progeny to contribute at a comparable level to other phenomena thought to be a source of such events, such as the high-energy components of cosmic rays.  Thus, such Rn-induced events may feature prominently as a persistent low-rate, high-impact phenomenon capable of circumventing gap-engineering efforts, and remain present as quantum computing scales to larger devices.  Furthermore, without careful mitigation, it will dominate the high-energy event rate in underground operation where hadronic components of cosmic ray backgrounds are heavily suppressed,

Over small-scale demonstrator chips, this plate-out occurs with sufficiently low rate that its effects are challenging to study directly. To overcome this, we have developed (\S\ref{sec:RadonPlateoutApparatus}) and characterized (\S\ref{sec:MeasurementsOfRadonDaughterPlateout}) a facility for enhanced radon plateout that can be used to controllably poison qubit devices and packaging with this background. We demonstrate a factor of $7\times10^4$ increase in plate-out rates relative to ambient plate-out\footnote{With respect to the prediction for our dead-air model under ambient conditions for the same exposure duration: 7 days.}, including the additional enhancement from applying an electric field. We have also validated a ``dead-air'' model of radon exposure (App.~\ref{appendix:deadair}) that reasonably reproduces the grow-in of radon daughters from exposures in this facility. Future work is aimed at depositing \(^{222}\)Rn daughters on superconducting qubit devices in this facility and measuring their impacts on qubit coherence metrics.

The dead-air model's predictions applied to example superconducting qubit devices, as discussed in Section~\ref{sec:projections}, suggest radon daughters as a potentially dominant source of multi-MeV particle impacts, but these predictions depend heavily on an assumed exposure history. This background can therefore be mitigated through several best-practices in fabrication, packaging, and installation/operation. Plateout-prevention techniques center around strictly limiting the integrated time over which a device \textit{and its enclosure} are exposed to ambient air. Strategies for such prevention include device staging in dry N\(_{2}\) purge boxes and purged bags, double-bagging devices and enclosures and backfilling with dry N\(_{2}\) for device transport, and the use of Mylar or Nylon bagging material, which demonstrate reduced diffusion radon diffusion constants compared to polyethylene bags~\cite{MengBagging,MylarBorexino,LinehanGridProduction}. Keeping device and enclosure exposures contained within cleanroom environments is expected to reduce the plate-out rate due to the potential of HEPA filters to capture \(^{222}\)Rn daughters produced outside of a cleanroom~\cite{LinehanLZRadon2026}. Moreover, dedicated radon-reduced cleanrooms which prevent not only ingress of \(^{222}\)Rn daughters but also \(^{222}\)Rn itself, may also be employed to further reduce plate-out. This strategy~\cite{Pushkin2018}, employed by the low-background dark matter community, has demonstrated ambient radon reduction by a factor of 2200~\cite{Akerib2020b}.

For devices that have already experienced a non-negligible amount of plateout, wiping and/or rinsing will reduce the attached radon fraction on a surface, but will not fully remove all daughters due to an embedded component, as we have shown in this work. However, chemical etching or electropolishing to remove the top few hundred nanometers of a surface can reduce this component, if such a process is possible for a given device or enclosure~\cite{hoppe2007cleaning,zuzel2012removal2,Schnee2014a, guiseppe2018review,bruenner2021radon,zuzel2012removal,Schnee2014a}. Ultimately, as with low-background rare event searches, a combination of prevention and post-plate-out removal may best limit this background as superconducting quantum computing platforms continue to scale.

\section*{Acknowledgments}

The authors wish to thank B. Zatschler for sharing the implantation profiles for $^{210}$Pb that were input into the Geant4 simulation, and J. Fowler and the NIST Quantum Sensors Division of Ref.~\cite{Fowler2024} for providing the cosmic ray spectra and associated processing script. This work was produced by the FermiForward Discovery Group, LLC under Contract No. 89243024CSC000002 with the U.S. Department of Energy (DOE), Office of Science (OS), Office of High Energy Physics (OHEP). This work was supported in part by the DOE including Grant No. DE-SC0014223, National Quantum Information Science Research Center: Quantum Science Center, and the DOE OS OHEP Program Office. Publisher acknowledges the U.S. Government license to provide public access under the DOE Public Access Plan. SSP performed the radon plateout on all samples and collected and analyzed all data from the alpha counter.  SSP calculated the relevant efficiencies and corrections. RL and DJT led the manuscript drafting, prepared the copper samples, fit and interpreted the data and designed the simulation studies. SSP contributed to manuscript drafting. DJT developed the dead-air model and ran the SRIM simulations. AR and MH developed and executed the \textsc{Geant4} simulation.  DK performed electrostatic simulations to determine vertical placement of the test samples in the exposure apparatus. NR performed the RAD7 detector measurements and analysis. DB, EFF, LH, and RS provided resources and support for this work, while LH provided scientific oversight.

\appendix

\section{Alpha Counter Geometric Efficiency} 
\label{appendix:alphacounter}

\begin{figure}[t]
\centering
\includegraphics[width=1.0\linewidth]{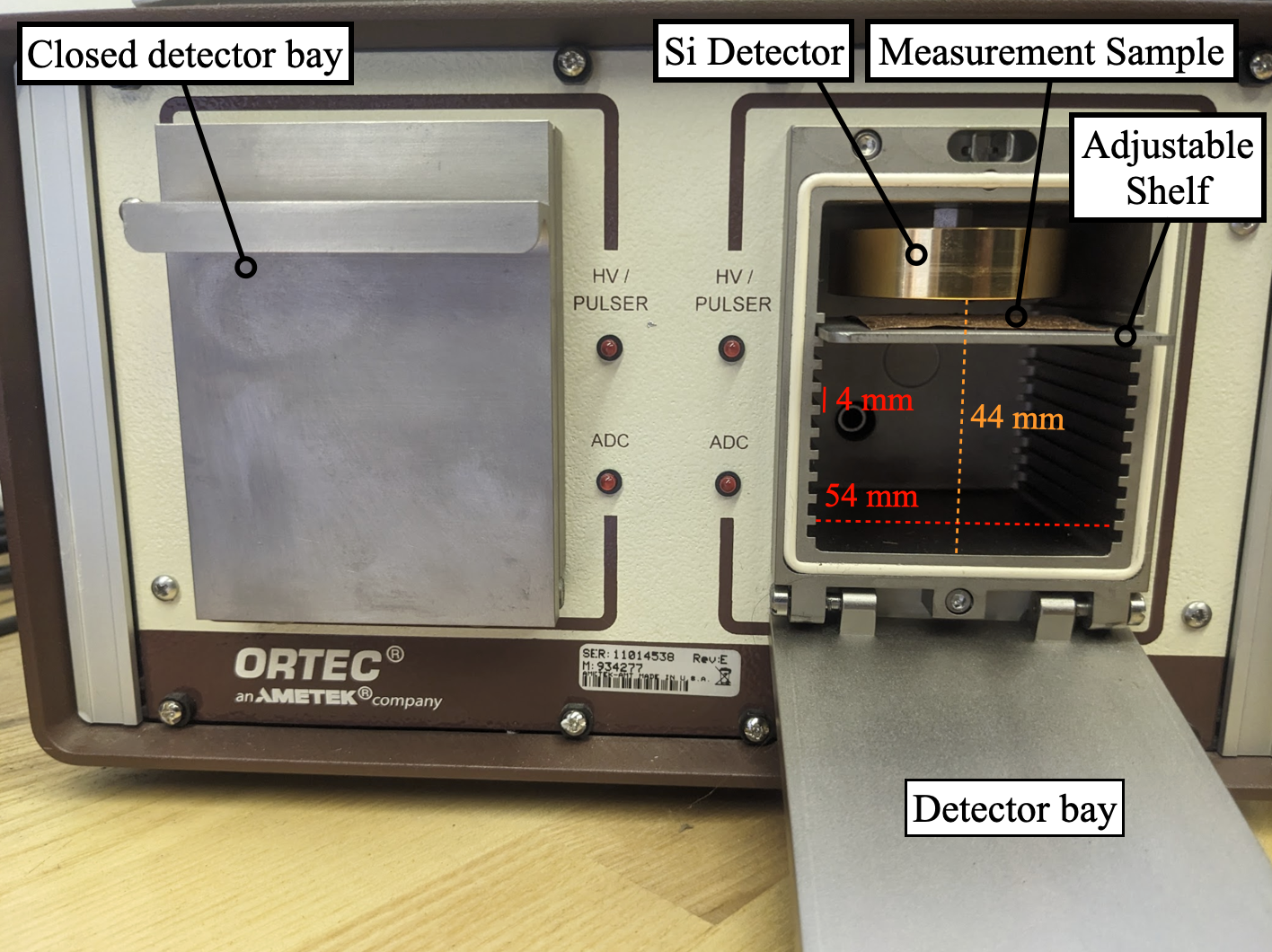}
\caption{Ortec Alpha Duo Alpha Spectrometer}
\label{fig:alphacounter}
\end{figure}

An Alpha Duo spectrometer~\cite{OrtecAlphaDuoManual} was used to measure $^{210}$Po alphas emitted by the Cu test pieces following the plateout. The Alpha Duo (shown in Figure~\ref{fig:alphacounter}) is a dual-chamber alpha spectrometer with pressure-regulated vacuum chambers for precise measurements of alpha activity in a sample. The adjustable sample shelves allow the user to position the sample between 4 and 44 mm below the detector. Each detector functions independently and offers fully adjustable energy ranges from 0 to 10 MeV. The bias supply is adjustable between 0 V and $\pm$100 V. In all measurements reported in this work, the bias voltage was set at 50 V. The silicon detectors used in the Alpha Duo are cylindrical with active detection disc of radius 1.69~cm and thickness 300~$\mu$m. Alphas in the energy range (1-10)~MeV deposit energy within the few 10s-of-$\mu$m in the silicon.

\begin{figure}[t]
\centering
\includegraphics[width=1.0\linewidth]{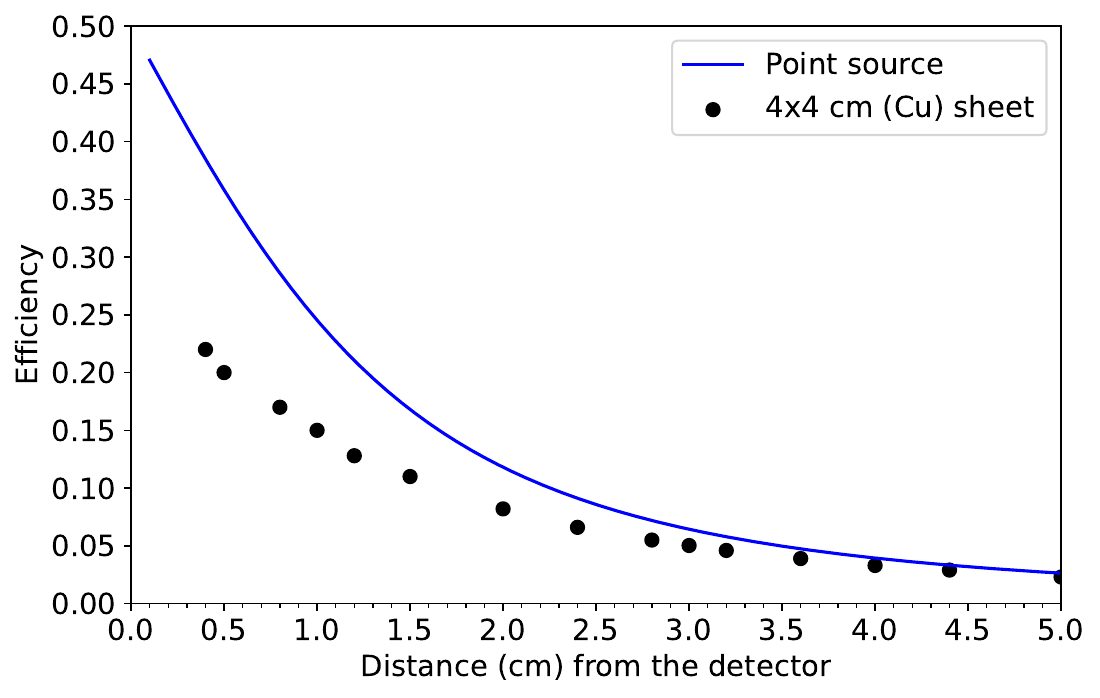}
\caption{The detection efficiency as a function of distance between sample and detector for a point source (analytic) and a $4\times 4$ cm$^2$ Cu foil (simulated). }
\label{fig:efficiency_curve}
\end{figure}

The detector efficiency and the energy deposition spectrum can be obtained using Monte Carlo simulations. We use FLUKA \cite{Ahdida2022,Battistoni2015} simulations to obtain the detector efficiency and energy deposition spectra. A simplified geometry is taken which consists of 4 cm x 4 cm Cu foil (of thickness 0.2 cm) placed directly below a Si detector with radius $R=1.69$ cm and thickness 2 cm. In the simulations, 5.304 MeV alphas were generated uniformly with isotropic initial momentum in the top 60 nm surface of the Cu foil. The fraction of events that deposit non-zero energy in the silicon detector active medium was determined and defined to be the efficiency. For the Cu foil placed 5 mm away, the efficiency is 20\%. In Fig.~\ref{fig:efficiency_curve}, we show the efficiency for a point source (using an analytic expression) and a 4$\times$4-cm$^2$ sheet as a function of the vertical distance between the sample and detector. 

\begin{table}[b]
\small
\begin{tabular}{|l|c|c|}
 \hline
 \textbf{Sample}& \textbf{Separation (mm)} & \textbf{Effective efficiency} \\
 \hline
 Cu coin & 3  & 41 $\%$  \\
 \hline 
 Cu foils & 4 & 22 $\%$\\
 \hline
  \end{tabular}
  \caption{Effective $^{210}$Po alpha geometric detection efficiency for each of the Cu test samples based on the Monte Carlo simulations. The separation column indicates the distance between top surface of the sample and the detector.}
  \label{tab:Effective_efficiencies}
\end{table}

The efficiencies used to convert the measured count rates into the $^{210}$Po activity for the measurements presented in the main text are tabulated in Table~\ref{tab:Effective_efficiencies}. Note these efficiencies assume uniform plateout (see \S\ref{subsec:ThinFoilTestsEField}).

\section{\textsc{Rad7} Measurements at High Activity} \label{app:rad7-saturation}
At radon specific activity $>$ 0.75 MBq/m$^3$ \cite{DurridgeRAD7Manual}, the \textsc{Rad7} measurements aren't reliable due to pile-up of events. Thus, when the \textsc{Rad7} reports radon concentrations of 4-7 MBq/m$^3$, we expect the actual $^{222}$Rn concentration to be larger. To quantify this underestimation, we perform a dedicated measurement to determine a correction factor at the high radon specific activities experienced in our facility. Durridge, the manufacturer of the \textsc{Rad7} indicates that such a measurement will apply for a given RAD7, although the resulting curve might be different for different units~\cite{DurridgePrivComms}.

We perform a single standard injection of the $^{222}$Rn, as described in the main text, and make measurements of the concentration every 2--3 days over the course of 3.4 weeks (6.3 half-lives of $^{222}$Rn) while the radon decays. These specific-activity measurements are then fit at late times, when the reported activity is $<0.75$ MBq/m$^3$, to an exponential decay. The obtained curve allows extrapolation to the specific activity at early times. 

\begin{figure}[t]
\centering
\includegraphics[width=1.0\linewidth]{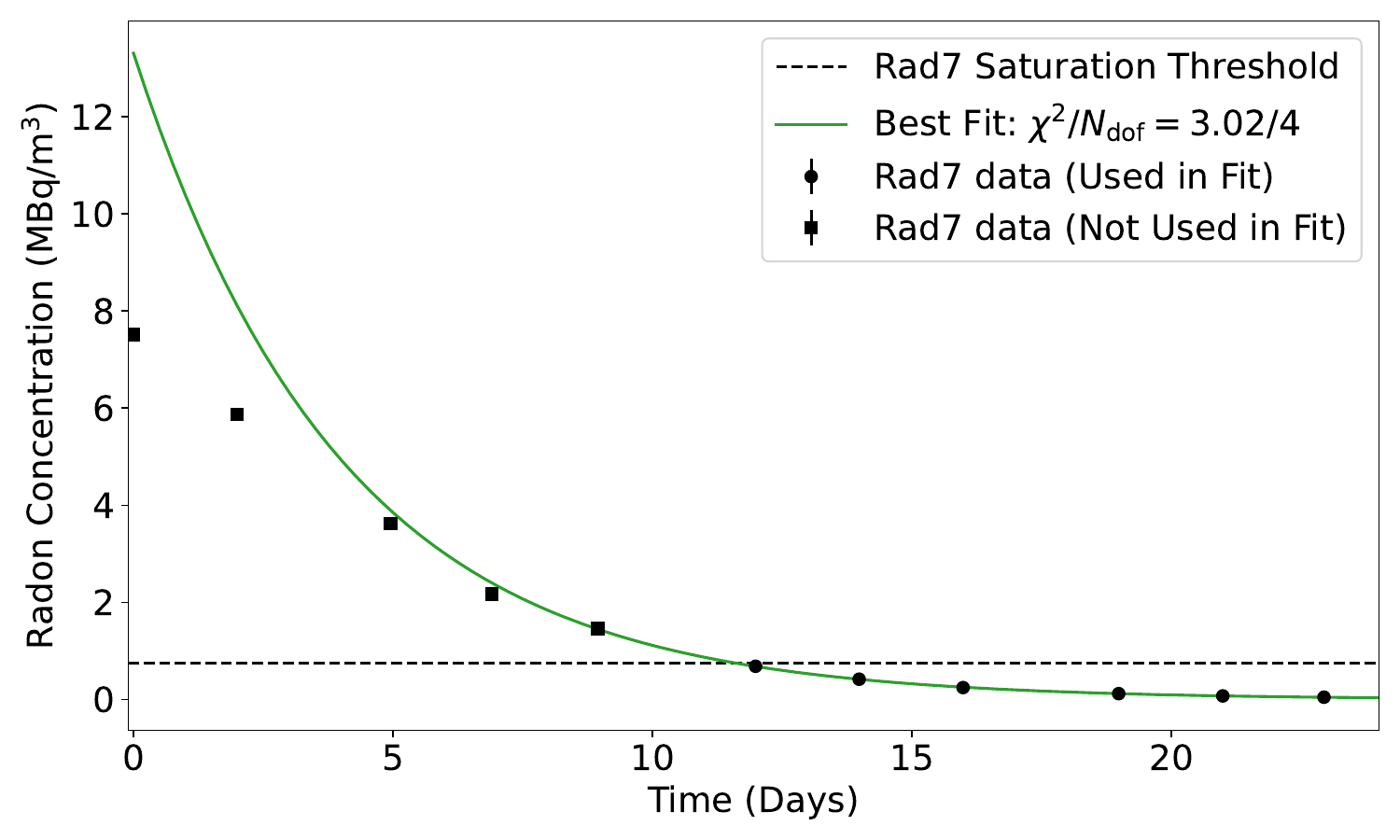}
\caption{An exponential fit to the \textsc{Rad7} data at late times. Only data points (circles), which are below 0.75 MBq/m$^3$ are used in the fit. The total $\lambda$ obtained corresponds to a half-life of 2.8 days, and includes both $\lambda_\mathrm{decay}$ and $\lambda_\mathrm{leak}$, suggesting effective leak rate of 0.067 day$^{-1}$. Uncertainties on all measured points are below 6\%.}
\label{fig:rad7-correction}
\end{figure}

The \textsc{Rad7} data, best-fit curve, and its extrapolation are shown in Fig.~\ref{fig:rad7-correction}. We expect some radon leakage, primarily from the \textsc{Rad7}, during these measurements; as a result, the total decay rate ($\lambda$) we obtain from the fit is larger than the radioactive decay constant for $^{222}$Rn. This discrepancy is attributed to outbound leaks during \textsc{Rad7} measurements, which we quantify using a model: 
\begin{align}
    \rho_A(t)=\rho_{A,0} e^{-(\lambda_\mathrm{Rn} + \lambda_\mathrm{leak})t}~,
\end{align}
where $\lambda_\mathrm{leak}$ is the effective leak rate which we find to be 0.067$\pm$0.003 day$^{-1}$ and $\lambda_{\mathrm{Rn}}$ is the natural $^{222}$Rn decay rate (fixed in fit). In order to perform this measurement, the volumes of the pressure cooker and the \textsc{Rad7} are shared, which results in a geometric correction of 1.2$\times$ the concentration as measured by the \textsc{Rad7}. This geometric correction is applied to the concentrations measured by the \textsc{Rad7} throughout this paper (\textit{e.g.}, the data in Fig.~\ref{fig:rad7-correction}). 

The \textsc{Rad7} saturation correction varies depending on the concentration measured: higher measured concentrations have larger saturation correction factors. We find, for example, when the \textsc{Rad7} reads an activity of $4.5$ MBq/m$^3$, the geometry correction factor (1.2) and saturation correction factor of 1.35 corresponds to a real specific activity of 7.3 MBq/m$^3$ initially injected into the pressure cooker for a 7 day build-up of radon in Pylon source.

\section{Dead-Air Model}
\label{appendix:deadair}

The dead-air model, used as a common point of reference for the plateout enhancement discussed in this work, and for projecting the impact of it on qubit chips, assumes the following: a volume $V$, with internal exposed surface area $A$ (initially free of radon progeny) is evacuated. At time $t=0$ it is instantaneously filled with an atmosphere of constant volume activity $\rho_A$ of $^{222}$Rn (specified in Bq/m$^3$), and is constantly replenished to maintain the constant activity (to represent the quasi-static emanation rate of radon). We do not include any effects from electric fields, convection, air flow, or gravity. Ultimately, we seek to understand the total activity of the radon progeny (per unit area of the volume's surface) after a time \(t\) of exposure.

In this scenario, the number density of $^{222}$Rn atoms, $n_\mathrm{Rn}$, is a constant determined by $\rho_A$:
\begin{align}
    n_\mathrm{Rn} = \frac{\rho_A}{\lambda_\mathrm{Rn}}~,
\end{align}
where $\lambda_\mathrm{Rn} = \log (2) / T_{1/2}^\mathrm{Rn}$ is the $^{222}$Rn decay constant given by its half-life $T_{1/2}^\mathrm{Rn}$. Note that due to the quasi-static emanation rate, the number density of Rn atoms is constant, thus the activity volume density is equal to the rate at which radon emanates into the volume (as for each atom that decays, another must be added to keep the activity constant). 

For each daughter nucleus, we must track the number density of atoms of that species in the gas and on the surface. To do so, we define, for the $i$-th daughter, $n_i(t)$ to be the number density in the gas (given in atoms/m$^3$) and $\sigma_i(t)$ to be the areal number density on the surface (given in atoms/m$^2$). The corresponding activity densities in the gas and on the surface are 
\begin{align}
    a_i^{(g)} &= \lambda_i n_i(t) \\
    a_i^{(s)} &= \lambda_i \sigma_i(t)~,
\end{align}
respectively. It is trivial to see that $a_{0} (t) \equiv \rho_A$. The condition that the gas and surface are initially free of progeny sets $n_i(t=0)=\sigma_i(t=0) = 0$.

To model the plate-out process, we define for each daughter, a plate-out velocity $v_i$ which determines the flux of atoms onto the surface: $J_i = n_i v_i$. This corresponds to a gas loss rate of $(dn_i / dt)_\mathrm{plateout} = -(A/V) J_i$. Thus, the total loss rate from decay and plate out of the $i$-th daughter is $\Lambda_i \equiv \lambda_i + (A/V) v_i$, which carries units of $\mathrm{s}^{-1}$. The differential equations governing the gas-phase number density are
\begin{align} \label{eq:ODE-gas-base}
    \frac{d n_i}{d t} = \lambda_{i-1} n_{i-1} - \Lambda_i n_i ~,
\end{align}
for $i \in [1,7]$ to reach $^{210}$Po, taking $\lambda_{0} n_{0} = \rho_A$. As the gas volume at $t=0$ is initially filled only with $^{222}$Rn, there are no progeny present. Thus we adopt the initial conditions $n_i(0)=0$. 

Likewise, the equations governing the evolution of surface densities are
\begin{align} \label{eq:ODE-surf-base}
    \frac{d \sigma_i}{d t} = \lambda_{i-1} \sigma_{i-1} + v_i n_i  - \lambda_i \sigma_i~,
\end{align}
with $\sigma_0(t) = 0$, as the radon itself is not plating out on the surface, but remains in the gas. Similarly, since the surface is initially free of any progeny, we take $\sigma_i(0)=0$. 

The resulting coupled differential equations are similar to the Bateman equations, but are augmented by the inclusion of deposition and removal terms that couple the gas and surface populations. Readers familiar with modeling radon plate out may note the similarity of our model to the Jacobi model~\cite{Jacobi1972, Porstendorfer1978}. The model presented here differs from the Jacobi formalism in several important respects. First, we neglect any attachment of progeny to airborne particulate, and any losses due to filtration or ventilation, corresponding to the case of dust-free stagnant air. In this limit, the only mechanism for loss of airborne progeny other than radioactive decay is deposition onto surfaces. Second, we do not enforce steady-state or secular equilibrium conditions. Instead, we retain the full time dependence of the coupled gas and surface populations, then solve the transient buildup dynamically. Secular equilibrium in the early chain (the short-lived components) emerges naturally. Finally, whereas the Jacobi model tracks separately the unattached and aerosol-attached progeny populations, our model combines these transport processes into an effective deposition velocity. We simplify our model by applying a single average deposition velocity $v_d$ to all progeny: $v_i = v_d$ (as motivated by Ref.~\cite{mishra2009measurement}). Thus, the quantity $\kappa \equiv (A/V)v_d$ is a constant (carrying dimensionality of inverse time) for every species.

For indoor conditions typical of those in a pressure vessel or a clean room (\textit{i.e.}, largely dust-free), the average deposition velocity of unattached radon progeny is $v_d = 7.9~\mathrm{m/hr}$~\cite{mishra2009measurement}, but can vary by factors up to a few~\cite{knutson1983radon} depending on room conditions such as air circulation and ventilation. Regardless, $v_d$ is fast compared to the lifetime of $^{210}$Pb. As such, it is a safe assumption that nearly all $^{210}$Pb atoms will find a surface onto which to plate out before decaying. Surfaces exposed to radon-laden air thus accumulate a surface activity of $^{210}$Pb from which an activity of $^{210}$Po grows-in.

Now, with these coupled differential equations that describe the linear, constant-coefficient, driven chain and our initial conditions, we can write the solutions for the number density of each daughter in the gas, parameterized in terms of its steady-state:
\begin{align} \label{eq:gas-closed-form-1}
    n_i(t) &= n_i(\infty) + \sum_{j=1}^i C_{ij} e^{-\Lambda_j t} \\  &= n_i(\infty) + C_{ii} e^{-\Lambda_i t} + \sum_{j=1}^{i-1} C_{ij} e^{-\Lambda_j t}~,\label{eq:gas-closed-form-1}
\end{align}
where $n_i(\infty)$ is found by setting the gas-phase ODEs to steady state and solving recursively:
\begin{align} \label{eq:gas-steady-state-sol}
    n_j(\infty) \equiv \lim_{t \to \infty} n_j(t) = \rho_A \frac{\prod_{m=1}^{j-1} \lambda_m}{\prod_{m=1}^{j} \Lambda_m}~.
\end{align}
For $j=1$, the numerator is an empty product which evaluates to unity, yielding:
\begin{align}
    n_1(\infty) = \frac{\rho_A}{\Lambda_1}~,\qquad n_{i>1}(\infty) = \frac{\lambda_{i-1}}{\Lambda_i} n_{i-1}(\infty)~.
\end{align}
By enforcing the boundary condition $n_1(0)=0$, we can compute our coefficients, starting with $i=j=1$:
\begin{align}
    C_{11} = -n_1(\infty) = -\frac{\rho_A}{\Lambda_1}~.
\end{align}
In the case of $j=i$:
\begin{align} \label{eq:diag-gas-const}
    C_{ii} = -n_i(\infty) - \sum_{j=1}^{i-1} C_{ij}~.
\end{align}

To determine $C_{i,j<i}$, we return to the ODE for $n_1(t)$ (Eq.~\ref{eq:ODE-gas-base}). We then insert the expression for $n_{i-1}(t)$ (Eq.~\ref{eq:gas-closed-form-1}) and multiply both sides by a factor $e^{\Lambda_i t}$. We then make a change of variable $t\to\tau$, and apply the operator $\int_0^t d \tau$ to both sides of the equation. Multiplying the resulting equation by $e^{-\Lambda_i t}$ on both sides and working through the algebra yields the following:
\begin{align*}
    n_i(t) =&~ n_i(\infty) \left[ 1 - e^{-\Lambda_i t} \right] \\ &~+ \sum_{j=1}^{i-1} C_{i-1,j} \frac{\lambda_{i-1}}{\Lambda_i - \Lambda_j} \left[ e^{-\Lambda_j t} - e^{-\Lambda_i t} \right]~,
\end{align*}
which we desire to match the form of Eq.~\ref{eq:gas-closed-form-1}. By matching terms with the coefficient $e^{-\Lambda_j t}$, we can read off
\begin{align}
    C_{ij} = \frac{\lambda_{i-1}}{\Lambda_i - \Lambda_j} C_{i-1,j}~,
\end{align}
for $j<i$. Similarly, from the terms with $e^{-\Lambda_i t}$, we recover Eq.~\ref{eq:diag-gas-const}. We now have a recursive formula to generate the time-dependent gas number density of each daughter isotope: $n_i(t)$.

The general solution for the number density of the $i$-th daughter on the surface is
\begin{align} \label{eq:sol-surf-density}
    \sigma_i(t) = \sigma_i(\infty) + \sum_{j=1}^{i} A_{ij} e^{-\Lambda_j t} + \sum_{j=1}^{i} B_{ij} e^{-\lambda_j t}~,
\end{align}
where $\sigma_i(\infty)$ is the steady-state solution, $A_{ij}$ and $B_{ij}$ are unknown coefficients, and assuming all the poles are distinct. 

We may determine $\sigma_i(\infty)$ by way of the steady-state solution for the gas number density steady-state solution for the $j$-th daughter. By evaluating the differential equations as $t\to\infty$, where the $d\sigma_i/dt\to0$, one arrives at
\begin{align}
    \sigma_i(\infty) = \frac{v_d}{\lambda_i} \sum_{j=1}^i n_j (\infty)~,
\end{align}
or
\begin{align}
    \sigma_i(\infty) = \frac{v_d}{\lambda_i} \rho_A \sum_{j=1}^i \frac{\prod_{m=1}^{j-1} \lambda_m}{\prod_{m=1}^{j} (\lambda_m + \kappa)}~.
\end{align}
Note that the surface activity density $a^\mathrm{(s)}_i(t\to\infty)$ is the product of the above and $\lambda_i$.

Now we are left to determine the coefficients $A_{ij}$ and $B_{ij}$ in Eq.~\ref{eq:sol-surf-density}. We proceed in a similar fashion: begin by decomposing the sums in Eq.~\ref{eq:sol-surf-density} into terms of $A_{ii}+B_{ii}$ and $A_{ij} + B_{ij}$ for $j<i$. We then evaluate the boundary condition $\sigma_1(0)=0$, from which we obtain
\begin{align}
    B_{11} = -\sigma_1 (\infty) - A_{11}~.
\end{align}
To obtain an expression for $A_{11}$, we proceed in a similar fashion as for the $C_{ij}$ coefficients. First, we insert the expression for $n_1(t)$ into Eq.~\ref{eq:ODE-surf-base} and apply the transfer-function method described above to obtain an expression for $\sigma_1(t)$ that can be compared to \begin{align}
    \sigma_1(t) = \sigma_1(\infty) + A_{11}e^{-\Lambda_1 t} + B_{11} e^{-\lambda_1 t}~,
\end{align}
allowing us to identify
\begin{align}
    \sigma_1(\infty) &= \frac{v_d}{\lambda_1} n_1(\infty) \\
    A_{11} &= \frac{v_d C_{11}}{\lambda_1 - \Lambda_1} = \frac{v_d}{\kappa} n_1(\infty)~.
\end{align}
For the $A_{ij}$ coefficients, we begin from the ODE for $\sigma_{i>1}(t)$, inserting the exponential expansions for the gas and surface components, and apply the derivative. We then again group terms by their exponential factors, and by inspection identify
\begin{align}
    A_{ij} &= \frac{\lambda_i A_{i-1,j}+ v_d C_{ij}}{\lambda_i-\Lambda_j}~~~\mathrm{for}~1\le j\le i\\
    B_{ij} &= \frac{\lambda_{i-1}}{\lambda_i - \lambda_j} B_{i-1,j}~~~\mathrm{for}~j<i~.
\end{align}
The coefficient $A_{ii}$ is a special case of the recursion above, in which we set $A_{i-1,j}=0$ since $\sigma_{i-1}$ does not contribute a pole at $\Lambda_i$. Finally, from the boundary conditions $\sigma_i(0)=0$, we obtain our constraint
\begin{align}
    B_{ii} &= -\sigma_i(\infty) - \sum_{j=1}^{i}A_{ij} - \sum_{j=1}^{i-1}B_{ij}~.
\end{align}
We now have a complete analytic solution for the dead-air model that can be generated recursively. 

For completeness, we explore the activity of the daughters after the exposure period, where the radon-laden atmosphere is evacuated from the volume. In this case, there is no source of new daughters, only those already on the surface contribute to the future activity. In this case, the evolution of the surface density of the $i$-th daughter is
\begin{align}
    \frac{d\sigma_i}{dt} = \lambda_{i-1} \sigma_{i-1} - \lambda_i \sigma_i~,
\end{align}
as $n_i(t)=0$. This takes the exact form of the gas density time evolution during the exposure, but with different boundary conditions: $\sigma_i(0)=\sigma_i^\mathrm{exp}$ and $\sigma_i(\infty)=0$, where $\sigma_i^\mathrm{exp}$ is the final surface number density of the $i$-th daughter at the end of the exposure period. Following the same procedure as for Eq.~\ref{eq:ODE-gas-base}, we write:
\begin{align}
    \sigma_i(t) = D_{ii}e^{-\lambda_i} + \sum_{j=1}^{i-1} D_{ij} e^{-\lambda_j t}~,
\end{align}
with
\begin{align}
    D_{11} &= \sigma_1^\mathrm{exp} \\
    D_{ij} &= \frac{\lambda_{i-1}}{\lambda_i - \lambda_j} D_{i-1,j}~~(\mathrm{for}~j<i)\\
    D_{ii} &= \sigma_i^\mathrm{exp} - \sum_{j=1}^{i-1} D_{ij} ~.
\end{align}

Considering only the late chain, ignoring all radon progeny before $^{210}$Pb, and making some simplifying assumptions, we can arrive at a compact, closed-form expression for our marker of interest: $^{210}$Po. The short half-lives of the early progeny motivates this approach, as those rapidly decay away to $^{210}$Pb. The simplifying assumptions are as follows: (1) The rapid decays of any early-chain progeny on the surface contribute only a modest inflation of the $\sigma_{^{210}\mathrm{Pb}}^\mathrm{exp}$, which we neglect; and (2) we also neglect the initial activity of the daughters of $^{210}$Pb: $\sigma_2^\mathrm{exp}=\sigma_3^\mathrm{exp}=0$. This scenario is equivalent to having a surface with some areal density of $^{210}$Pb atoms and no other components of the chain.

Under these simplifying assumptions, the surface activity density is
\begin{widetext}
\begin{align} 
    a_\mathrm{Po}^{(s)}(t) =& \lambda_\mathrm{Po} \sigma_\mathrm{Po}(t)\\ 
    =& ~\sigma_\mathrm{Pb}^\mathrm{exp} \lambda_\mathrm{Pb} \lambda_\mathrm{Bi} \lambda_\mathrm{Po} \left[\frac{e^{-\lambda_\mathrm{Pb}t}}{(\lambda_\mathrm{Bi}-\lambda_\mathrm{Pb})(\lambda_\mathrm{Po}-\lambda_\mathrm{Pb})} +
        \frac{e^{-\lambda_\mathrm{Bi}t}}{(\lambda_\mathrm{Pb}-\lambda_\mathrm{Bi})(\lambda_\mathrm{Po}-\lambda_\mathrm{Bi})} +
        \frac{e^{-\lambda_\mathrm{Po}t}}{(\lambda_\mathrm{Pb}-\lambda_\mathrm{Po})(\lambda_\mathrm{Bi}-\lambda_\mathrm{Po})} \right],
        \label{eq:finalBatemaneq}
\end{align}
\end{widetext}
which we use throughout this work to project the grow-in of $^{210}$Po. Note in the above equation the subscripts all refer to isotopes of mass 210. While at short times after the exposure end, the simplified form above underestimated the activity of $^{210}$Po, at long times Eq.~\ref{eq:finalBatemaneq} asymptotes to the value predicted by the full analytic solution. Thus, the $\mathcal{A}_\mathrm{3y}$ predicted by the full solution and by Eq.~\ref{eq:finalBatemaneq} are equivalent (provided $t_\mathrm{exp} < 3$ years).

\begin{figure}[t]
\includegraphics[width=1.0\linewidth]{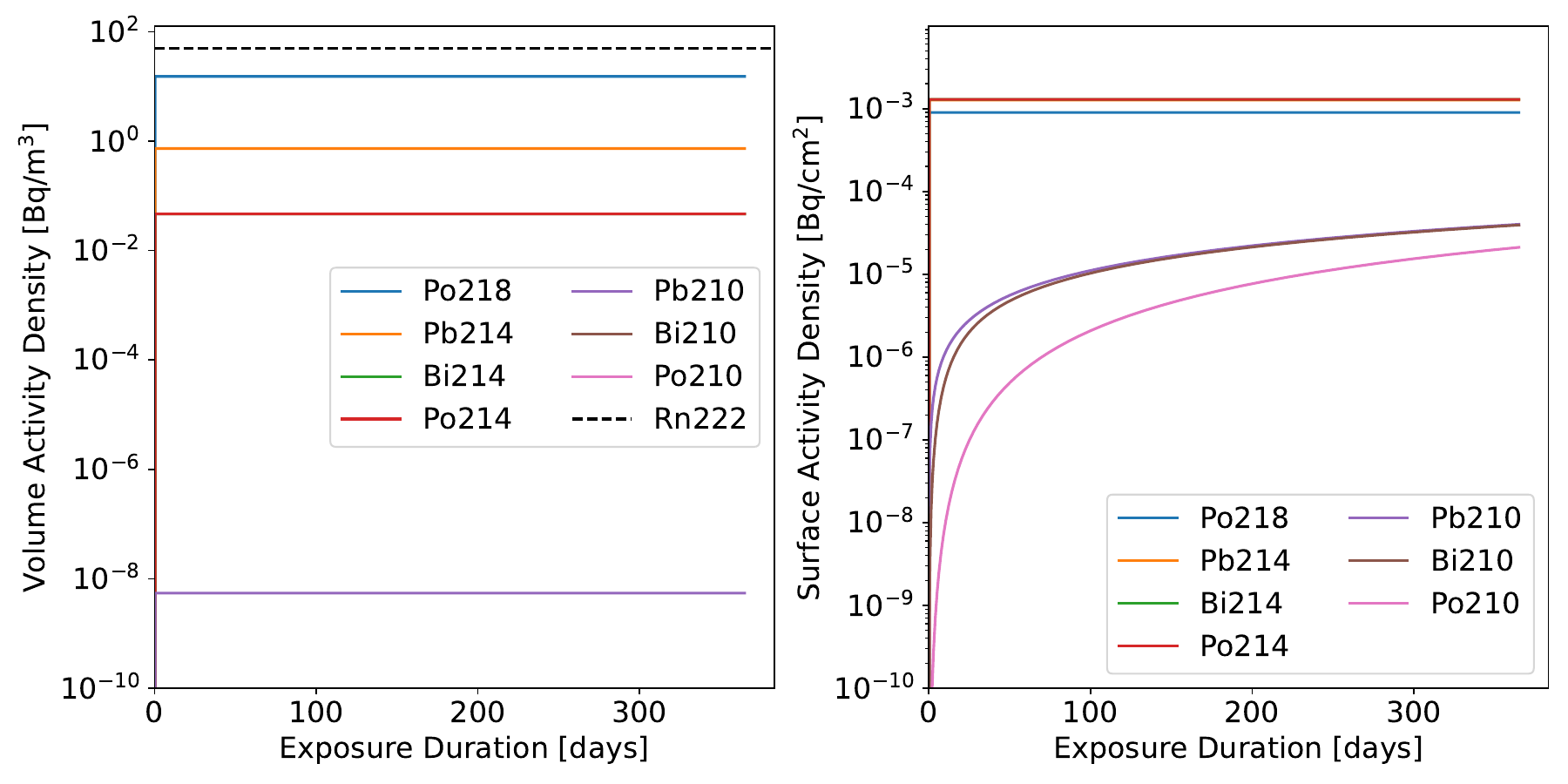} \\
\includegraphics[width=1.0\linewidth]{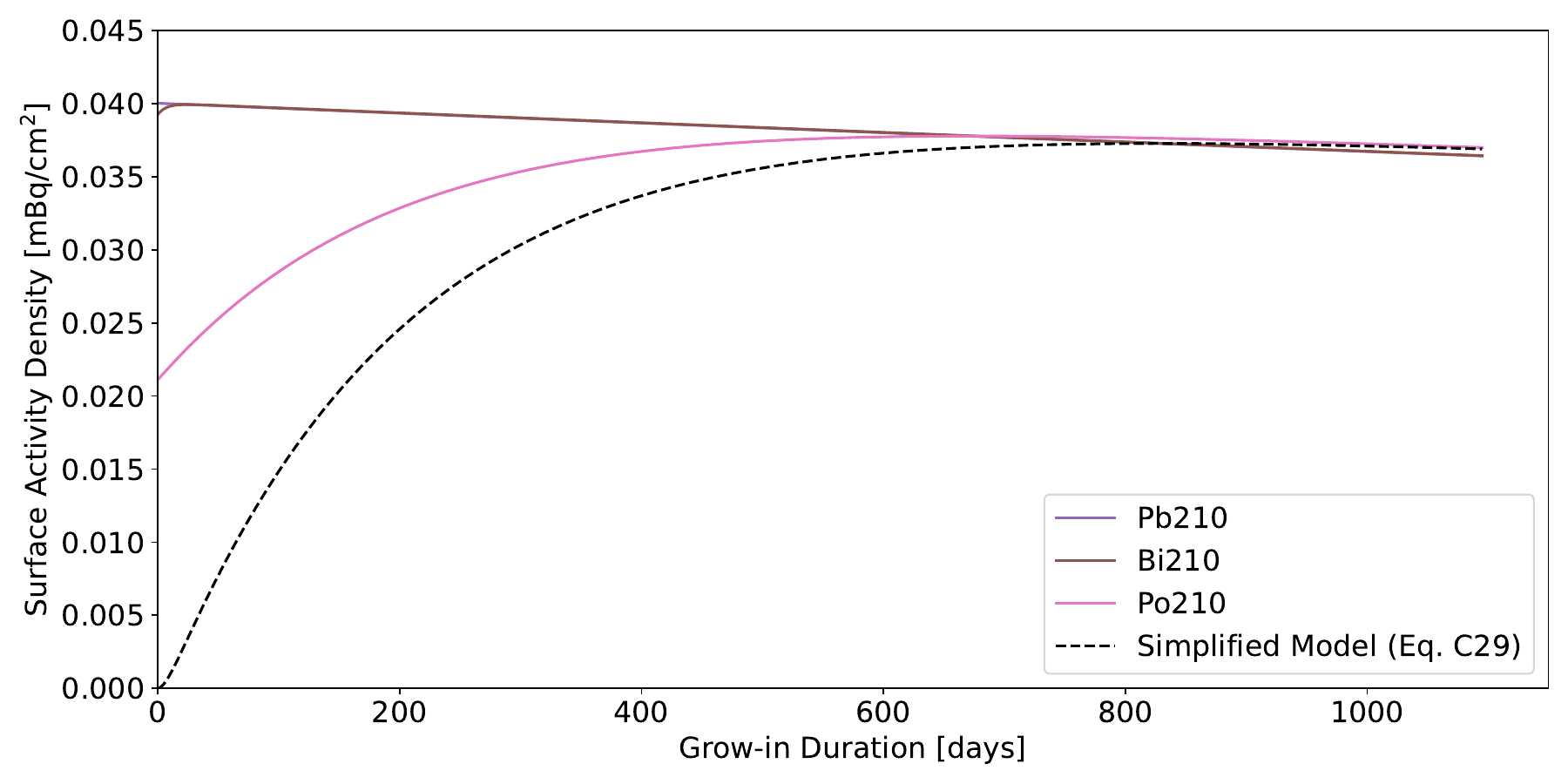}
\caption{Evolution of the Dead-Air model during the exposure period (top) and grow-in period (bottom) for ambient conditions ($\rho_A=50$ Bq/m$^3$, $\ell=1.55$ m). During the exposure period, volume (top left) and surface (top right) activity densities are shown for each daughter in the $^{222}$Rn chain. For the grow-in period, only the late-chain daughters are shown. Also shown with the dashed line in the grow-in plot is the simplified equation for $a^{(s)}_\mathrm{Po}$ (Eq.~\ref{eq:finalBatemaneq}).} 
\label{fig:dead-air-curves}
\end{figure}

We explore this model for the ``ambient" case of $\rho_A=50$ Bq/m$^3$ with $v_d=7.9$ m/hr~\cite{mishra2009measurement}. We choose to parameterize our area-to-volume ratio as $A/V=6/\ell$, where $\ell$ is the vessel's characteristic length scale\footnote{This parameterization holds for many simple shapes. For a cube, $\ell$ is the side length; for a sphere, the diameter. It holds also for a cylinder with a unity aspect ratio: $2r=h=\ell$. However, when considering a room-like setting with furniture, the presence of these features inflates $A/V$ (as each object adds surface area and reduces the total volume containing radon). This effectively serves as a reduced value of $\ell$ in this parameterization.}, and choose 
$\ell=1.55$ m ($\kappa=1.4\times10^{-4}$ s$^{-1}$). The $^{210}$Pb surface activity achieved with one year of exposure is $a_\mathrm{Pb}^{(s)}=0.04$~mBq/cm$^2$, (about 1 decay every six hours per cm$^2$),
with $a_\mathrm{Po}^{(s)}$ about a factor of two lower. The $^{210}$Po surface activity after three years of grow-in is $\mathcal{A}_\mathrm{3y}^\mathrm{model}=0.037$ mBq/cm$^2$. While this length scale may appear unrepresentative of a fabrication or laboratory setting, we note that this model does not account for ventilation. We choose this length scale to be roughly consistent with the $^{210}$Pb deposition rate as measured at SNOLAB~\cite{stein2018radon}, corrected for the difference in measured radon concentration. Our model predicts 111.6 $^{210}$Po atoms deposited per day under these conditions. The exposure and grow-in of each daughter's activity under this scenario is shown in Fig.~\ref{fig:dead-air-curves}

\begin{figure}[b]
\includegraphics[width=1.0\linewidth]{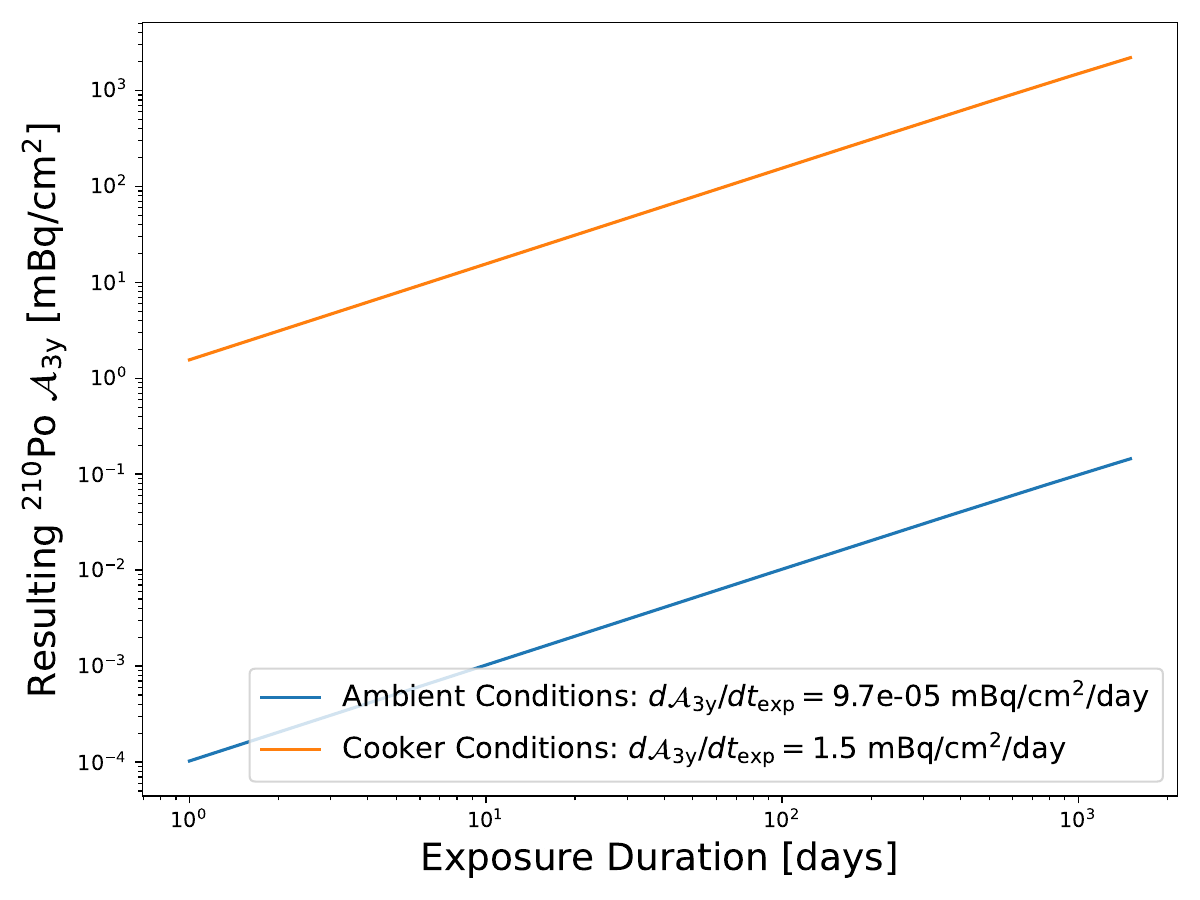}
\caption{Increase in $\mathcal{A}_\mathrm{3y}$ as a function of exposure duration, as predicted by the dead-air model, for the two scenarios considered here: ``ambient" ($\rho_A=$50 Bq/m$^3$, $\ell=1.55$m) and ``cooker" ($\rho_A=$7.5 MBq/m$^3$, $A/V=38.2$ m$^{-1}$).} 
\label{fig:dead-air-A3y}
\end{figure}

For the plate-out apparatus discussed in this work ($A/V=38.2~\mathrm{m}^{-1}$ and $\rho_A=7.3~\mathrm{MBq/m}^3$), for a seven-day exposure, we obtain $a_\mathrm{Pb}^{(s)}=11.4$ mBq/cm$^2$ at the end of the exposure, corresponding to $\mathcal{A}_\mathrm{3y}^\mathrm{model}=10.6$~mBq/cm$^2$ of $^{210}$Po. We measure an $\mathcal{A}_\mathrm{3y}\approx13$~mBq/cm$^2$, see \S\ref{subsec:ThinFoilTestsNoField}), for these cooker conditions (no applied field). In Fig.~\ref{fig:dead-air-A3y}, we show the increase in $\mathcal{A}_\mathrm{3y}$ as a function of exposure time for both this scenario and the ambient scenario described above.

Also note that this model can be extended to a decaying radon concentration ($\rho_A \to \rho_A(t) = \rho_{A,0} e^{-\lambda_0 t}$) with the same solutions but removing the steady state terms and extending the pole set for the $C_{ij}$ and $A_{ij}$ coefficients to include $\lambda_0$, the $^{222}$Rn decay constant. \\

\section{\textsc{Geant4} Simulations}
\label{appendix:g4-sims}

To produce the background estimates in Fig.~\ref{fig:G4-spectra}, we simulated decays of \(^{210}\)Pb, \(^{210}\)Bi, and \(^{210}\)Po within a benchmark chip-plus-enclosure volume in Geant4. We simulate a silicon chip with dimensions $5 \times 5 \times 0.5$ mm$^3$ (to match the dimensions of the geometry in Ref.~\cite{Fowler2024}), centered within a cavity formed from a copper enclousure with inner dimensions $5.2 \times 5.2 \times 1.5$ mm$^3$ and wall thickness 15.4~mm. In between the chip and enclosure we simulate vacuum. The physics of the simulation is primarily governed by four \textsc{Geant4} physics lists. \textsc{G4RadioactiveDecayPhysics} governs the decay of isotopes, \textsc{G4EmStandardPhysics\_option4} governs the set of electromagnetic interactions in the simulation, while the response of recoiling heavy ions is governed by the \textsc{G4ScreenedRecoils} and \textsc{G4LindhardPartition} physics lists. 

Decays of $^{210}$Pb, $^{210}$Bi, and $^{210}$Po are generated uniformly in-plane across the surfaces of both the copper enclosure and silicon chip. As a fraction of these decays may come from daughters embedded into the surface (\S\ref{sec:RadonDaughterPlateoutBackgrounds}) we spatially localize the depth of these decays relative to the surface following one of three distributions. The first distribution arises if \(^{218}\)Po is the first daughter to land on the surface, and is modeled as a truncated gaussian profile. The parameters defining this depth distribution are different for copper and silicon, and are obtained from simulations performed in Ref.~\cite{BirgitPrivateComm}. If a daughter after \(^{218}\)Po but before (and including) \(^{214}\)Po is the first to land on the surface, this yields the second distribution, another truncated gaussian. Finally, we also include a purely surface component, which we simulate as planar source on face of the chip or enclosure. As estimating the relative contributions of these three components is nontrivial to do \textit{a-priori}, the simulations in Fig.~\ref{fig:G4-spectra} arbitrarily assume equal weighting between the three. This assumption is likely to only minimally bias our overall results, because only the \(^{206}\)Pb recoil peak from \(^{210}\)Po decay experiences substantial attenuation from embedding.

The \(^{210}\)Pb, \(^{210}\)Bi, and \(^{210}\)Po isotope decays were then simulated, with $10^7$ events of each isotope simulated from the copper housing and $10^6$ events of each isotope simulated from the chip. For each event, a total in-chip energy was extracted; these energies were binned to create a spectrum for a given isotope and origin (chip or enclosure). We define an efficiency factor by dividing the number of events which produce an energy deposition in the silicon by the total number of events simulated. Each resulting spectrum is rescaled such that its integrated rate, corrected with the efficiency factor, matches a target surface activity. This target activity is determined by the dead-air model prediction for each isotope's surface activity density three years after the end of the radon exposure\footnote{This corresponds to the rates on the right hand side of Fig.~\ref{fig:dead-air-curves} (bottom), and yields to nearly identical scalings for each isotope.}.  We then sum the spectra for each isotope of interest independently for events originating in the enclosure and originating in the silicon. An energy-dependent smoothing function (identical to that used in Ref.~\cite{Fowler2024}) is then applied to the data. The resulting scaled curves are as presented in Fig.~\ref{fig:G4-spectra}.

\bibliographystyle{unsrt}
\bibliography{main-abridged.bib}

\end{document}